\documentclass[journal]{IEEEtran}
\usepackage{amsmath, amssymb, amsthm, mathtools, graphicx, subfigure, paralist, relsize, hyperref}

\newtheorem{theorem}{Theorem}

\newtheorem{corollary}[theorem]{Corollary}
\newtheorem{lemma}{Lemma}
\newtheorem{conjecture}{Conjecture}
\newtheorem{definition}{Definition}
\newtheorem{example}{Example}
\newtheorem{remark}{Remark}

\newcommand*{\ldblbrace}{\{\mskip-6mu\{}
\newcommand*{\rdblbrace}{\}\mskip-6mu\}}
\newcommand{ \C }{ \mathcal C }
\newcommand{ \alf }[1]{ [#1] }
\newcommand{ \Lat }{ \mathcal L }
\newcommand{ \da }{ d_{\textnormal{a}} }

\newcommand{ \sml }[1]{ \mathsmaller{#1} }

\newcommand{ \cube }{ \operatorname{cub} }
\newcommand{ \conv }{ \operatorname{con} }
\newcommand{ \ceil }[1]{ \left\lceil #1 \right\rceil }
\newcommand{ \floor }[1]{ \left\lfloor #1 \right\rfloor }

\newcommand{ \vol }{ \operatorname{Vol} }
\newcommand{ \myqed }{\hfill $\blacktriangle$}
\newcommand{ \defeq }{ \coloneqq }

\begin{document}

\title{Codes in the Space of Multisets---\\Coding for Permutation Channels with Impairments}

\author{
        Mladen~Kova\v{c}evi\'{c} and Vincent~Y.~F.~Tan%
\thanks{Date: January 11, 2018.
        %Manuscript received July 10, 2017; revised November 23, 2017; accepted December 22, 2017.
				
				This work was supported by the Singapore Ministry of Education (MoE) Tier 2 grant
        ``Network Communication with Synchronization Errors: Fundamental Limits and Codes'' (Grant number R-263-000-B61-112).
				Part of the work was presented at the 2017 IEEE International Symposium on Information Theory (ISIT) \cite{kovacevic+tan_isit}.
        
				The authors are with the Department of Electrical and Computer Engineering,
        National University of Singapore, Singapore 117583
				(emails: \{mladen.kovacevic, vtan\}@nus.edu.sg).				
				V. Y. F. Tan is also with the Department of Mathematics,
				National University of Singapore, Singapore 119076.
				%
				%Communicated by V. Sidorenko, Associate Editor for Coding Theory.
				%
				%DOI 10.1109/TIT.2017.2789292
}%
}

\maketitle

\begin{abstract}
  Motivated by communication channels in which the transmitted sequences are subject
to random permutations, as well as by certain DNA storage systems, we study the error
control problem in settings where the information is stored/transmitted in the form
of multisets of symbols from a given finite alphabet.
A general channel model is assumed in which the transmitted multisets are potentially
impaired by insertions, deletions, substitutions, and erasures of symbols.
Several constructions of error-correcting codes for this channel are described, and
bounds on the size of optimal codes correcting any given number of errors derived.
The construction based on the notion of Sidon sets in finite Abelian groups is shown
to be optimal, in the sense of the asymptotic scaling of code redundancy, for any
``error radius'' and any alphabet size.
It is also shown to be optimal in the stronger sense of maximal code cardinality in
various cases.
\end{abstract}


\section{Introduction}
\label{sec:introduction}

\IEEEPARstart{I}{nformation} storage systems using pools of DNA molecules
as storage media have recently been proposed in the literature \cite{heckel}.
A distinctive feature of those models is that the data is ``written'' in the
pools in an unordered fashion.
This kind of storage is rather different from the traditional ones, and requires
information to be encoded in the form of \emph{multisets}%
\footnote{Informally, a multiset is a set with repetitions of elements allowed.}
of symbols---objects which are unordered by definition.
A similar situation arises in communication channels in which the input sequences
are subject to random permutations.
In such channels, the order of the symbols belonging to the input sequence cannot
be inferred by the receiver with a reasonable degree of confidence, and the only
carrier of information is again the multiset of the transmitted symbols.

One of the necessary ingredients of both kinds of systems mentioned above are
codes capable of protecting the stored/transmitted multisets from various types
of noise.
Motivated by this observation, we study error-correcting codes in the space of
multisets over an arbitrary finite alphabet.
The error model that we consider is ``worst-case'' (as opposed to probabilistic);
in other words, we focus on constructions and bounds on the cardinality of
optimal codes capable of correcting a given number of errors.

\subsection{The channel model}
\label{sec:channel}

We next describe the channel model that will be referred to throughout the paper.
More concrete examples of communication channels that served as motivation for
introducing this model are mentioned in the following subsection.

The channel inputs are multisets of symbols from a finite alphabet $ \mathbb{A} $
that we shall, without loss of generality, assume to be $ [q] \defeq \{0, 1, \ldots, q-1\} $, for some $ q \geq 2 $.
Let $ U = \ldblbrace u_1, \ldots, u_n \rdblbrace $ denote%
\footnote{We use a double-braces notation for multisets;
e.g., $ \ldblbrace a,a,b \rdblbrace $ denotes a multiset containing two copies
of $ a $ and one copy of $ b $.}
the generic input, where
$ u_i \in \mathbb{A} $ and the $ u_i $'s are not necessarily all distinct.
The channel acts on the transmitted multiset $ U $ by removing some of its elements
(deletions), by adding some elements to it (insertions), by replacing some of its
elements with other symbols from $ \mathbb{A} $ (substitutions), and by replacing
some of its elements with the symbol ``$ ? $'' (erasures).
As a result, another multiset
$ \tilde{U} = \ldblbrace \tilde{u}_1, \ldots, \tilde{u}_{\tilde n} \rdblbrace $ is
obtained at the channel output, where $ \tilde{u}_i \in \mathbb{A} \cup \{?\} $ and
$ \tilde{n} = |\tilde{U}| $ in general need not equal $ n = |U| $.
The goal of the receiver is to reconstruct $ U $ from $ \tilde{U} $.

As pointed out above, our main object of study are codes enabling the receiver
to uniquely recover the transmitted multiset $ U $ whenever the total number of
errors that occurred in the channel is smaller than some specified threshold.

\subsection{Motivation}
\label{sec:motivation}

\subsubsection{Permutation Channels}

Consider a communication channel that acts on the transmitted sequences by permuting
their symbols in a random fashion.
In symbolic notation:
\begin{equation*}
  u_1\ u_2\ \cdots\ u_n  \quad  \rightsquigarrow  \quad  u_{\pi(1)}\ u_{\pi(2)}\ \cdots\ u_{\pi(n)} ,
\end{equation*}
where $ (u_1, \ldots, u_n) \in \mathbb{A}^n $ is a \emph{sequence} of symbols from the
input alphabet $ \mathbb{A} $, and $ \pi $ is drawn uniformly at random from the set of
all permutations over $ \{1, \ldots, n\} $.
Apart from shuffling their symbols, the channel is  assumed to impose other kinds of
impairments on the transmitted sequences as well, such as insertions, deletions,
substitutions, and erasures of symbols.
We refer to this model as the \emph{permutation channel}%
\footnote{To the best of our knowledge, apart from \cite{kovacevic+vukobratovic_desi},
there is only a handful of papers discussing coding and related problems for channels
with random reordering of symbols, e.g., \cite{walsh, metzner, gadouleau1, kovacevic+vukobratovic_clet}.
Channel models with restricted reordering errors have also been studied in the literature,
e.g., \cite{ahlswede+kaspi, krachkovsky, langberg, schulman};
in these and similar works it is assumed that only certain permutations are admissible during
a given transmission and, consequently, they require quite different methods of analysis.
}
\cite{kovacevic+vukobratovic_desi}.

As discussed in \cite{kovacevic+vukobratovic_desi}, the appropriate space for defining
error-correcting codes for the permutation channel is the set of all \emph{multisets}
over the channel alphabet $ \mathbb{A} $, and therefore the results of this paper are
relevant precisely for such models.
The reasoning behind this observation is very simple: in the permutation channel,
no information can be transferred in the \emph{order} of the transmitted symbols
because this information is irretrievably lost due to random shuffling.
Consequently, the only carrier of information is the multiset of the transmitted
symbols $ \ldblbrace u_1, \ldots, u_n \rdblbrace $.

The communication scenario that motivated introducing the permutation channel
model are packet networks employing multipath routing as a means for end-to-end
packet transfer \cite{kurose+ross}.
In such networks, packets belonging to the same ``generation'' usually traverse
paths of different lengths, bandwidths, congestion levels, etc., on their way to the
receiver, which causes their delays to be different and unpredictable.
Consequently, the packets may arrive at the destination in a different order than the
one they were transmitted in.
Furthermore, packets can be deleted in the network due to buffer overflows in
network routers, link failures, etc., and they can also experience other types of errors
for various reasons.
Therefore, this can be seen as an instance of the permutation channel whose alphabet
is the set of all possible packets.

Models related to the permutation channel are also relevant for diffusion-based
molecular communications \cite{nakano} where reordering errors, as well as deletions,
are frequent due to the physical properties of the molecular transmission mechanisms.

\subsubsection{DNA Storage Systems}

A class of data storage systems that uses synthesized DNA molecules as information
carriers was recently proposed%
\footnote{A related model is also discussed in an unpublished manuscript \cite{mackay},
albeit from a different perspective.}
and studied in \cite{heckel}.
In this model, information is written onto $ n $ DNA molecules of length $ \ell $ each,
which are then stored in an unordered way.
In other words, information is stored in the form of multisets of cardinality $ n $ over
an alphabet of size $ 4^\ell $.

Two assumptions that were made in \cite{heckel} that make the model therein different
from ours are the following:
\begin{inparaenum}
\item[(i)]
the molecules from the stored multiset are sampled with replacement by the receiver, and
\item[(ii)]
no errors are introduced at the molecule level during the reading process.
\end{inparaenum}
Furthermore, the main problem addressed in \cite{heckel} are the fundamental limits of
the system in terms of rate and  under a vanishing error probability formalism.
Here, we are mostly concerned with the fundamental limits of the code sizes in the
space of multisets when various types of impairments are present in the channel.

We should note that most of the works on DNA storage assume that coding is performed
at the molecule level, with molecules regarded as quaternary \emph{sequences};
see \cite{yazdi}.
To avoid possible confusion we emphasize once more that what we are discussing in this
paper is error correction for a different model, partially motivated by \cite{heckel},
which assumes that codewords are \emph{multisets} of DNA molecules (or multisets of any
other objects for that matter).

\subsection{Main results and paper organization}

In Section \ref{sec:general} we introduce multiset codes formally and demonstrate
some of their basic properties.
We prove that all four types of impairments considered in this paper are in a sense
equivalent, so one can focus on analyzing only one of them, e.g., deletions.
We also introduce a metric that is appropriate for the problem at hand and that
characterizes the error correction capability of multiset codes.

A geometric restatement of the problem, given in Section \ref{sec:geometry}, reveals
a close connection between multiset codes and codes in the so-called $ A_m $ lattices
\cite{conway+sloane}, which prompted us to investigate the latter in their own right.
These results, presented in Section \ref{sec:An}, will be used subsequently to
derive some properties of multiset codes, but are also of independent interest.
In particular, we demonstrate that linear codes in $ A_m $ lattices are geometric
analogs of the so-called Sidon sets, a notion well-known in additive combinatorics
\cite{obryant}.
Perfect and diameter-perfect codes in $ A_m $ lattices are also studied here, and
several (non-)existence results in this context provided.

In Section \ref{sec:bounds} we describe our main construction of multiset codes,
based on Sidon sets, and derive bounds on the size of optimal codes correcting a
given number of errors.
This construction is shown to be optimal, in the sense of minimal asymptotic redundancy,
for any ``error radius'' and any alphabet size.
It is also shown to be optimal in the (stricter) sense of maximal code cardinality
in various cases.
It turns out that codes in the space of multisets are closely related to codes for
classical binary insertion/deletion channels with restrictions either on the noise
model, or on the channel inputs.
We discuss this fact in Section \ref{sec:indel} and show that our results improve
upon the existing results for those channels.

In Section \ref{sec:otherconstr} we describe two additional code constructions that are
provably suboptimal, but are of interest nonetheless.
One of them is based on indexing---a method that essentially turns a sequence into a set
by adding a sequence number prefix to each of its symbols.
This approach is frequently used to protect the packets from possible reorderings in
networking applications \cite{kurose+ross}.
We prove that indexing is strictly suboptimal in some regimes of interest and quantify
this fact by comparing the rates achievable by optimal codes obtained in this way to
those achievable by optimal multiset codes.
The other construction provided in Section \ref{sec:otherconstr} has the algebraic flavor
usually encountered in coding theory, and is based on encoding information in the roots
of a suitably defined polynomial.

A brief conclusion and several pointers for further work are stated in Section \ref{sec:conclusion}.

\section{General properties of multiset codes}
\label{sec:general}

As noted in Section \ref{sec:channel}, we shall assume throughout the paper that
the channel input alphabet is
$ \mathbb{A} = \alf{q} \defeq \{0, 1, \ldots, q-1\} $, for some $ q \geq 2 $.

\subsection{Basic definitions and geometric representation}
\label{sec:geometry}

For a multiset $ U = \ldblbrace u_1, \ldots, u_n \rdblbrace $ over $ \alf{q} $, denote
by $ {\bf x}^{U} = \left( x^{U}_0, x^{U}_1, \ldots, x^{U}_{q-1} \right) \in \mathbb{Z}^{q} $
the corresponding vector of multiplicities of its elements, meaning that $ x^{U}_i $ is
the number of occurrences of the symbol $ i \in \alf{q} $ in $ U $.
The vector $ {\bf x}^{U} $ satisfies $ x^{U}_i \geq 0 $ and $ \sum_{i=0}^{q-1} x^{U}_i = |U| = n $.
Multisets over a given alphabet are uniquely specified by their multiplicity vectors.
For that reason we shall mostly use the vector notation and terminology in the sequel,
occasionally referring to multisets;
it should be clear that these are just two different ways of expressing the same notion.

We emphasize that, even though the codewords will be described by integer vectors,
these vectors are not actually sent through the channel described in Section \ref{sec:channel}.
Namely, if $ {\bf x} = (x_0, x_1, \ldots, x_{q-1}) \in \mathbb{Z}^{q} $ is a codeword,
then what is being transmitted is a multiset containing $ x_0 $ copies of the symbol
$ 0 $, $ x_1 $ copies of the symbol $ 1 $, etc.
Therefore, $ n = \sum_{i=0}^{q-1} x_i $---the cardinality of the multiset in question---%
will be referred to as the \emph{length} of the codeword $ \bf x $ in this setting
because this is the number of symbols that are actually being transmitted.
Also, a deletion in the channel is understood not as a deletion of an element of $ \bf x $,
but rather a deletion of an element of the multiset represented by $ \bf x $, and
similarly for the other types of errors.

If we impose the usual requirement that all codewords are of the same length $ n $,
we end up with the following code space:
\begin{equation}
  \triangle_n^{q-1} = \left\{ {\bf x} \in \mathbb{Z}^{q}  :  x_i \geq 0,  \sum_{i=0}^{q-1} x_i = n \right\} .
\end{equation}
The set $ \triangle^{q-1}_n $ is a discrete simplex of ``sidelength'' $ n $,
dimension $ q-1 $, and cardinality $ |\triangle^{q-1}_n| = \binom{n + q - 1}{q - 1} $.

\begin{definition}
  A multiset code of length $ n $ over the alphabet $ \alf{q} $ is a subset of
$ \triangle_n^{q-1} $ having at least two elements.
\myqed
\end{definition}

The requirement that the code has at least two codewords is clear from the
practical perspective.
It is made explicit here to avoid discussing trivial cases in the sequel.

\begin{figure}[h]
 \centering
  \includegraphics[width=\columnwidth]{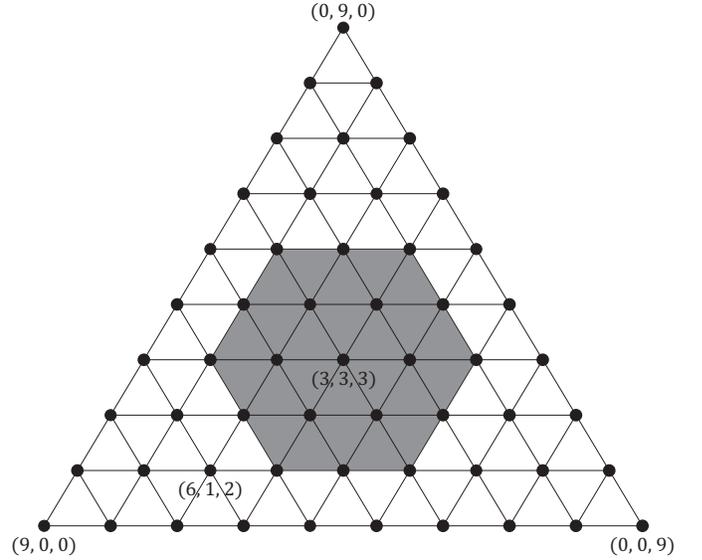}
\caption{The graph of the simplex $ \triangle^{2}_9 $ representing the set of all multisets
         of cardinality $ 9 $ over the ternary alphabet $ \{0, 1, 2\} $, and an illustration
				 of a ball of radius $ 2 $ in this graph.%
				}%
\label{fig:simplex}
\end{figure}

The metric on $ \triangle_n^{q-1} $ that is appropriate for our purposes is
essentially the $ \ell_1 $ distance:
\begin{equation}
\label{eq:metric}
  d_1({\bf x}, {\bf y}) = \frac{1}{2} \sum_{i=0}^{q-1} | x_i - y_i | .
\end{equation}
The metric space $ (\triangle_n^{q-1}, d_1) $ can be visualized as a graph with
$ |\triangle^{q-1}_n| = \binom{n + q - 1}{q - 1} $ vertices and with edges
joining vertices at distance one; see Figure \ref{fig:simplex}.
The quantity $ d_1({\bf x}, {\bf y}) $ is precisely the graph distance between the
vertices corresponding to $ \bf x $ and $ \bf y $, i.e., the length of the shortest
path between them.
The minimum distance of a code $ \C \subseteq \triangle_n^{q-1} $ with
respect to $ d_1(\cdot,\cdot) $ is denoted $ d_1(\C) $.

The following definition is motivated by the structure of the code space $ \triangle_n^{q-1} $.
Namely, this space can be seen as the translated $ A_{q-1} $ lattice%
\footnote{A lattice \cite{conway+sloane} in $ \mathbb{Z}^q $ is a set of vectors
$ {\mathcal L} \subseteq \mathbb{Z}^q $ that is closed under addition and subtraction.
In other words, $ ({\mathcal L},+) $ is a subgroup of $ (\mathbb{Z}^q,+) $.}
restricted to the non-negative orthant, where
\begin{equation}
\label{eq:An}
  A_{q-1} = \left\{ {\bf x} \in \mathbb{Z}^{q}  :  \sum_{i=0}^{q-1} x_i = 0 \right\} .
\end{equation}
Written differently, $ \triangle^{q-1}_n = (A_{q-1} + {\bf t}) \cap \mathbb{Z}^q_+ $,
where $ {\bf t} \in \mathbb{Z}^q $ is any vector satisfying $ \sum_{i=0}^{q-1} t_i = n $,
and $ \mathbb{Z}_+ \defeq \{0,1,\ldots\} $.

\begin{definition}
  We say that a multiset code $ \C \subseteq \triangle^{q-1}_n $ is linear if
$ \C = (\Lat + {\bf t}) \cap \triangle^{q-1}_n $ for some \emph{lattice}
$ \Lat \subseteq A_{q-1} $ and some vector $ {\bf t} \in \mathbb{Z}^{q} $ with
$ \sum_{i=0}^{q-1} t_i = n $.
\myqed
\end{definition}

In other words, $ \C $ is linear if it is obtained by translating a linear
code in $ A_{q-1} $ (a sublattice of $ A_{q-1} $) and keeping only the codewords
with non-negative coordinates.

\subsection{Error correction capability of multiset codes}

Let $ {\bf e}_i \in \mathbb{Z}^{q} $, $ i \in \alf{q} $, be the unit vector
having a $ 1 $ at the $ i $'th coordinate and $ 0 $'s elsewhere.
Let $ U $ be the transmitted multiset.
If the received multiset $ \tilde{U} $ is produced by inserting a symbol
$ i $ to $ U $, then
$ {\bf x}^{\tilde{U}} = {\bf x}^{U} + {\bf e}_i $.
Similarly, deletion of $ i $ from $ U $ means that
$ {\bf x}^{\tilde{U}} = {\bf x}^{U} - {\bf e}_i $,
and a substitution of $ i \in U $ by $ j $ that
$ {\bf x}^{\tilde{U}} = {\bf x}^{U} - {\bf e}_i + {\bf e}_j $.

We say that a code can correct $ h_{\text{ins}} $ insertions, $ h_{\text{del}} $
deletions, and $ h_{\text{sub}} $ substitutions if no two distinct codewords can
produce the same channel output after being impaired by \emph{arbitrary patterns}
of $ \leq h_{\text{ins}} $ insertions, $ \leq h_{\text{del}} $ deletions, and
$ \leq h_{\text{sub}} $ substitutions.
In other words, every codeword can be uniquely recovered after being impaired by
such an error pattern.

\begin{theorem}
\label{thm:indelsub}
  Let $ \C \subseteq \triangle_n^{q-1} $ be a multiset code,
$ h_{\textnormal{ins}} $, $ h_{\textnormal{del}} $, $ h_{\textnormal{sub}} $ non-negative
integers, and
$ h = h_{\textnormal{ins}} + h_{\textnormal{del}} + 2 h_{\textnormal{sub}} $.
The following statements are equivalent:
\begin{itemize}
\item[(a)]
$ \C $ can correct $ h_{\textnormal{ins}} $ insertions, $ h_{\textnormal{del}} $
deletions, and $ h_{\textnormal{sub}} $ substitutions,
\item[(b)]
$ \C $ can correct $ h $ insertions,
\item[(c)]
$ \C $ can correct $ h $ deletions.
\end{itemize}
\end{theorem}
\begin{IEEEproof}
  Since any substitution can be thought of as a combination of a deletion and an
insertion, and vice versa, the statement {\it (a)} is equivalent to the following:
\begin{itemize}
\item[{\it (a')}]
$ \C $ can correct $ h_{\text{ins}} + h_{\text{sub}} $ insertions and
$ h_{\text{del}} + h_{\text{sub}} $ deletions.
\end{itemize}
We show next that {\it (a')} $ \Rightarrow $ {\it (c)};
the remaining implications can be proven in a similar way.
We shall assume that $ n \geq h $, the statement otherwise being vacuously true.

Suppose that {\it (c)} does not hold.
This means that there are two different codewords (multisets) that can produce
the same channel output after $ h $ elements have been deleted from both of them%
\footnote{A code can correct up to $ h $ deletions if and only if it can correct
\emph{exactly} $ \min\{h, n\} $ deletions (meaning that exactly $ \min\{h, n\} $
symbols of the transmitted multiset are being deleted in the channel).}.
In other words, there exist $ {\bf x}, {\bf y} \in \C $, $ {\bf x} \neq {\bf y} $,
such that $ {\bf x} - {\bf f} = {\bf y} - {\bf g} $ for some vectors
$ {\bf f}, {\bf g} $ with $ f_i, g_i \geq 0 $,
$ \sum_{i=0}^{q-1} f_i = \sum_{i=0}^{q-1} g_i = h $
($ {\bf f} $ and $ {\bf g} $ represent patterns of $ h $ deletions from
$ \bf x $ and $ \bf y $, respectively).
Write $ {\bf f} = {\bf f}^{\text{del}} + {\bf f}^{\text{ins}} $ and
$ {\bf g} = {\bf g}^{\text{del}} + {\bf g}^{\text{ins}} $, where
$ {\bf f}^{\text{del}}, {\bf f}^{\text{ins}}, {\bf g}^{\text{del}}, {\bf g}^{\text{ins}} $
are arbitrary vectors satisfying
$ f^{\text{del}}_i, f^{\text{ins}}_i, g^{\text{del}}_i, g^{\text{ins}}_i \geq 0 $,
$ \sum_{i=0}^{q-1} f^{\text{del}}_i = \sum_{i=0}^{q-1} g^{\text{del}}_i = h_{\text{del}} + h_{\text{sub}} $,
$ \sum_{i=0}^{q-1} f^{\text{ins}}_i = \sum_{i=0}^{q-1} g^{\text{ins}}_i = h_{\text{ins}} + h_{\text{sub}} $.
Then\linebreak $ {\bf x} - {\bf f}^{\text{del}} + {\bf g}^{\text{ins}} = {\bf y} - {\bf g}^{\text{del}} + {\bf f}^{\text{ins}} $,
which means that $ \C $ cannot correct $ h_{\text{ins}} + h_{\text{sub}} $
insertions and $ h_{\text{del}} + h_{\text{sub}} $ deletions.
Hence, {\it (a'}) does not hold.
\end{IEEEproof}

\begin{remark}%[Erasures]
\label{rem:erasures}
\textnormal{
  Erasures can be included in the model too, but we have chosen not to do
so here because it would slightly complicate notation (due to the additional
symbol ``$ ? $'' in the output alphabet).
Namely, in the same way as in the above proof one can show that erasures are as
damaging as deletions:
A code $ \C \subseteq \triangle_n^{q-1} $ can correct $ h $ erasures if and only if
it can correct $ h $ deletions.
This is intuitively clear.
Namely, in the usual scenarios where information is encoded in the form of
\emph{sequences} of symbols, erasures are easier to deal with than deletions
because the receiver can see the \emph{positions} of the erased symbols.
In the case of multisets, however, positions are irrelevant and carry no information.
}

\textnormal{
We emphasize that the above statement about the equivalence of erasures and deletions
in the context of multisets is only true for codes whose codewords are all of the same
length, i.e., codes in $ \triangle_n^{q-1} $.
In the case of variable-length codes, which we do not analyze here, erased symbols
can reveal some information about the cardinality of the transmitted multiset to
the receiver, unlike deleted symbols which do not appear at the channel output.
}
\myqed
\end{remark}

In light of Theorem \ref{thm:indelsub} and Remark \ref{rem:erasures}, we may
assume that deletions are the only type of noise in the channel.

The following statement gives a metric characterization of the error correction
capability of a multiset code $ \C $.

\begin{theorem}
\label{thm:corrdel}
  A multiset code $ \C \subseteq \triangle_n^{q-1} $ can correct $ h $
deletions if and only if its minimum distance satisfies $ d_1(\C) > h $.
\end{theorem}
\begin{IEEEproof}
  Let $ {\bf x}, {\bf y} $ be two codewords at distance $ d_1(\C) $.
Then $ {\bf f} = {\bf x} - {\bf y} $ satisfies $ \sum_{i=0}^{q-1} f_i = 0 $ and
$ \sum_{i=0}^{q-1} |f_i| = 2 d_1(\C) $; see \eqref{eq:metric}.
Let $ {\bf f}^+ = \max\{{\bf f}, {\bf 0}\} $ and $ {\bf f}^- = \max\{-{\bf f}, {\bf 0}\} $
be the positive and the negative part of $ \bf f $, respectively, so that
$ {\bf f} = {\bf f}^+ - {\bf f}^- $ (here $ \max $ is the coordinate-wise maximum).
Then $ {\bf x} - {\bf f}^+ = {\bf y} - {\bf f}^- $.
Since $ f^+_i, f^-_i \geq 0 $ and $ \sum_{i=0}^{q-1} f^+_i = \sum_{i=0}^{q-1} f^-_i = d_1(\C) $,
both $ {\bf f}^+$ and $ {\bf f}^- $ can be thought of as noise vectors describing
patterns of $ d_1(\C) $ deletions from $ \bf x $ and $ \bf y $, respectively.
This means that $ \C $ cannot correct $ d_1(\C) $ deletions.
Reversing the argument, one sees that $ \C $ can always correct
$ < d_1(\C) $ deletions because assuming otherwise would imply
that there exist two codewords at distance $ < d_1(\C) $, which
is a contradiction.
\end{IEEEproof}

\subsection{Error detection capability of multiset codes}

We now briefly discuss the error \emph{detection} problem for the studied channel
and show that it admits a metric characterization similar to the one obtained for
error correction.

We say that a code can detect $ h_{\text{ins}} $ insertions, $ h_{\text{del}} $
deletions, $ h_{\text{sub}} $ substitutions, and $ h_{\text{ers}} $ erasures
if no codeword $ \bf x $ can produce another codeword $ {\bf y} \neq {\bf x} $
at the channel output after being impaired by an \emph{arbitrary pattern} of
$ \leq h_{\text{ins}} $ insertions, $ \leq h_{\text{del}} $ deletions, $ \leq h_{\text{sub}} $
substitutions, and $ \leq h_{\text{ers}} $ erasures.
In other words, any such error pattern results in either the transmitted codeword
$ {\bf x} $, or something which is not a codeword at all, meaning that the receiver
can decide with certainty whether an error has happened during transmission or not.

Erasures are trivial to detect.
Also, if the number of insertions that occur in the channel is different from
the number of deletions, the received multiset will have a different cardinality
than the transmitted one and the detection is easy.
If the number of insertions and deletions is the same, say $ s $, then this can be
thought of as $ s $ substitutions, as discussed before.
Therefore, for the purpose of analyzing error detection, it is not a loss of
generality to consider substitutions as the only type of noise in the channel.

\begin{theorem}
  A multiset code $ \C \subseteq \triangle_n^{q-1} $ can detect $ h $
substitutions if and only if its minimum distance satisfies $ d_1(\C) > h $.
\end{theorem}

In other words, a code $ \C \subseteq \triangle_n^{q-1} $ can detect
$ h $ substitutions if and only if it can correct $ h $ deletions.
\vspace{2mm}

\begin{IEEEproof}
  That $ \C $ \emph{cannot} detect $ h $ substitutions means that there
are two different codewords $ {\bf x}, {\bf y} $, and a vector $ \bf f $ with
$ \sum_{i=0}^{q-1} f_i = 0 $, $ \sum_{i=0}^{q-1} |f_i| \leq 2h $, such that
$ {\bf y} = {\bf x} + {\bf f} $\linebreak
($ \bf f $ represents a pattern of $ h $ substitutions).
If this is the case, then $ d_1({\bf x}, {\bf y}) \leq h $, and hence $ d_1(\C) \leq h $.
The other direction is similar.
\end{IEEEproof}

\section{Codes in $ A_m $ lattices}
\label{sec:An}

As we observed in Section \ref{sec:geometry}, the space in which multiset codes
over a $ q $-ary alphabet are defined is a translated $ A_{q-1} $ lattice restricted
to the non-negative orthant.
This restriction is the reason why the space $ \triangle_n^{q-1} $ lacks some properties
that are usually exploited when studying bounds on codes, packing problems, and the like.
In order to analyze the underlying geometric problem we shall disregard these
constraints in this section and investigate the corresponding problems in the
metric space $ (A_{q-1}, d_1) $.
In particular, we shall discuss constructions of codes in $ A_{q-1} $ lattices having
a given minimum distance, bounds on optimal codes, and (non-)existence of perfect
and diameter-perfect codes in $ (A_{q-1}, d_1) $.
These results will be used in Section \ref{sec:bounds} to study the corresponding
questions for multiset codes, but are also of independent interest.

For notational convenience, throughout this section we denote the dimension of the
space by $ m $, rather than $ q - 1 $.

\subsection{$ A_{m} $ lattice under $ \ell_1 $ metric}

We first state some properties of $ A_m $ lattices under the metric $ d_1(\cdot, \cdot) $
defined in \eqref{eq:metric}.
As in the case of multiset codes, the minimum distance of a code $ \C \subseteq A_m $
with respect to the metric $ d_1(\cdot, \cdot) $ is denoted $ d_1(\C) $.
A code $ \C \subseteq A_m $ is said to be linear if it is a sublattice of $ A_m $.
For $ S, \C \subseteq A_m $, both nonempty, we say that $ (S, \C) $ is a \emph{packing}
in $ A_m $ if the translates $ S + {\bf x} $ and $ S + {\bf y} $ are disjoint for every
$ {\bf x}, {\bf y} \in \C $, $ {\bf x} \neq {\bf y} $.
If $ \C $ is a lattice, such a packing is called a lattice packing.
The definitions for $ \mathbb{Z}^m $ in place of $ A_m $, and for an arbitrary metric
in place of $ d_1(\cdot, \cdot) $, are similar.

Another way of describing codes in the metric space $ (A_m, d_1) $ will be convenient
for our purpose.
For $ {\bf x} = (x_1, \ldots, x_m),  {\bf y} = (y_1, \ldots, y_m) \in \mathbb{Z}^m $,
define the metric
\begin{equation}
  \da({\bf x}, {\bf y}) \defeq
    \max \left\{ \sum_{\substack{ i = 1 \\ x_i > y_i}}^{m} (x_i - y_i) ,
                 \sum_{\substack{ i = 1 \\ x_i < y_i}}^{m} (y_i - x_i) \right \} .
\end{equation}
This metric is used in the theory of codes for asymmetric channels (hence the
subscript `a'); see \cite[Ch.\ 2.3 and 9.1]{klove}.

\begin{theorem}
\label{thm:isometry}
  $ (A_m, d_1) $ is isometric to $ (\mathbb{Z}^m, \da) $.
\end{theorem}
\begin{IEEEproof}
  For $ {\bf x} = (x_0, x_1, \ldots, x_m) $, denote $ {\bf x}' = (x_1, \ldots, x_m) $.
The mapping $ {\bf x} \mapsto {\bf x}' $ is the desired isometry.
Just observe that, for $ {\bf x}, {\bf y} \in A_m $,
\begin{equation}
  d_1({\bf x}, {\bf y}) = \sum_{\substack{ i = 0 \\ x_i > y_i}}^{m}  (x_i - y_i)
                        = \sum_{\substack{ i = 0 \\ x_i < y_i}}^{m}  (y_i - x_i)
\end{equation}
because $ \sum_{i=0}^{m} x_i = \sum_{i=0}^{m} y_i = 0 $.
Then, by examining the cases $ x_0 \lessgtr y_0 $, it follows that
\begin{equation}
\begin{aligned}
  d_1({\bf x}, {\bf y}) 
    &=  \max \left\{ \sum_{\substack{ i = 1 \\ x_i > y_i}}^{m}  (x_i - y_i) , 
                     \sum_{\substack{ i = 1 \\ x_i < y_i}}^{m}  (y_i - x_i) \right\}  \\
    &=  \da({\bf x}', {\bf y}') .
\end{aligned}
\end{equation}
Furthermore, the mapping $ {\bf x} \mapsto {\bf x}' $ is bijective.
\end{IEEEproof}

Therefore, packing and similar problems in $ (A_m, d_1) $ are equivalent
to those in $ (\mathbb{Z}^m, \da) $, and hence we shall use these metric
spaces interchangeably in the sequel.
When discussing packings in $ (\mathbb{Z}^m, \da) $, the following sets
naturally arise:
\begin{equation}
\label{eq:Sn}
  S_m(r^\sml{+}, r^\sml{-})  \defeq
   \left\{ {\bf x} \in \mathbb{Z}^m :
       \sum_{\substack{ i = 1 \\ x_i > 0}}^{m}  x_i \leq r^\sml{+} ,
       \sum_{\substack{ i = 1 \\ x_i < 0}}^{m}  | x_i | \leq r^\sml{-} \right\} ,
\end{equation}
where $ r^\sml{+}, r^\sml{-} \geq 0 $.
$ S_m(r^\sml{+}, r^\sml{-}) $ is an anticode \cite{ahlswede} of maximum distance
$ r^\sml{+} + r^\sml{-} $ in $ (\mathbb{Z}^m, \da) $, i.e., a subset of
$ \mathbb{Z}^m $ of diameter $ r^\sml{+} + r^\sml{-} $.
In particular, $ S_m(r,r) $ is the ball of radius $ r $ around $ \bf 0 $ in
$ (\mathbb{Z}^m, \da) $, and $ S_m(r, 0) $ is the simplex---the set of all
non-negative vectors in $ \mathbb{Z}^m $ with coordinates summing to $ \leq r $.

\begin{lemma}
\label{thm:Anball}
 The cardinality of the anticode $ S_m(r^\sml{+}, r^\sml{-}) $ is
\begin{equation}
\label{eq:Anball}
  | S_m(r^\sml{+}, r^\sml{-}) | =
     \sum_{ j = 0 }^{ \min\{ m, r^\sml{+} \} }  \binom{m}{j}
                                                \binom{r^\sml{+}}{j}
                                                \binom{r^\sml{-} + m - j}{m - j} .
\end{equation}
\end{lemma}
\begin{IEEEproof}
  The $ j $'th summand in \eqref{eq:Anball} counts the vectors in
$ S_m(r^\sml{+}, r^\sml{-}) $ having $ j $ strictly positive coordinates.
These coordinates can be chosen in $ \binom{m}{j} $ ways.
For each choice, the ``mass'' $ \leq r^\sml{+} $ can be distributed over them
in $ \sum_{t=j}^{r^\sml{+}} \binom{t - 1}{j-1} = \binom{r^\sml{+}}{j} $ ways
(think of placing $ t \leq r^\sml{+} $ balls into $ j $ bins, where at least
one ball is required in each bin).
Similarly, the mass $ \leq r^\sml{-} $ can be distributed over the remaining
coordinates in
$ \sum_{t=0}^{r^\sml{-}} \binom{t+m-j-1}{m-j-1} = \binom{r^\sml{-} + m-j}{m-j} $ ways.
\end{IEEEproof}
\vspace{2mm}

The following claim provides a characterization of codes in $ (\mathbb{Z}^m, \da) $
in terms of the anticodes $ S_m(r^\sml{+}, r^\sml{-}) $.
We omit the proof as it is analogous to the proof of the corresponding statement
for finite spaces represented by distance-regular graphs \cite{delsarte, ahlswede}.

\begin{theorem}
\label{thm:codesaspackings}
  Let $ \C \subseteq \mathbb{Z}^m $ be a code, and $ r^\sml{+}, r^\sml{-} $ non-negative
integers.
Then $ (S_m(r^\sml{+}, r^\sml{-}), \C) $ is a packing if and only if
$ \da(\C) > r^\sml{+} + r^\sml{-} $.
\hfill \IEEEQED
\end{theorem}

Hence, whether $ (S_m(r^\sml{+}, r^\sml{-}), \C) $ is a packing depends on the
values $ r^\sml{+}, r^\sml{-} $ only through their sum.

\subsection{Codes in $ (\mathbb{Z}^m, \da) $: An upper bound}
\label{sec:Bh}

Since the space $ (\mathbb{Z}^{m}, \da) $ is infinite, we cannot use the cardinality
of a code as a measure of ``how well it fills the space''.
The infinite-space notion that captures this fact is the \emph{density} \cite{fejes-toth}
of a code, defined as
\begin{equation}
  \mu(\C)  \defeq  \lim_{k \to \infty}
    \frac{ \left| \C \cap \{-k, \ldots, k\}^{m} \right| }{ (2k+1)^{m} } .
\end{equation}
In case the above limit does not exist, one can define the upper ($ \overline{\mu}(\C) $)
and the lower ($ \underline{\mu}(\C) $) density by replacing $ \lim $
with $ \limsup $ and $ \liminf $, respectively.
For a linear code $ \C $ the density $ \mu(\C) $ exists and is equal to
$ \mu(\C) = \frac{1}{|\mathbb{Z}^{m} / \C|} $, where $ \mathbb{Z}^{m} / \C $
is the quotient group of the code/lattice $ \C $.
Clearly, the higher the required minimum distance, the lower the achievable density.
The following theorem quantifies this fact.

\begin{theorem}
\label{thm:boundsZq}
Let $ \C $ be a code in $ (\mathbb{Z}^{m}, \da) $ with minimum
distance $ \da(\C) = d $.
Then, for all non-negative integers $ r^\sml{+}, r^\sml{-} $ with
$ r^\sml{+} + r^\sml{-} < d $,
\begin{equation}
\label{eq:mu_upper}
  \overline{\mu}(\C)
							 \leq  | S_{m}(r^\sml{+}, r^\sml{-}) |^{-1} .
\end{equation}
In particular, for $ 2 \leq d \leq 2m + 1 $,
\begin{equation}
\label{eq:phik}
  \overline{\mu}(\C)
        <   \ceil{\frac{d-1}{2}}! \floor{\frac{d-1}{2}}!  \left( m + 1 - \ceil{\frac{d-1}{2}} \right)^{1-d}  ,
\end{equation}
and, for $ 1 \leq m < d $,
\begin{equation}
\label{eq:phih}
  \overline{\mu}(\C)  <  2^m m!^3 (2m)!^{-1} (d - m)^{-m} .
\end{equation}
\end{theorem}
In words, the theorem gives upper bounds on the density of codes in $ (\mathbb{Z}^m, \da) $
having a given minimum distance $ d $ and dimension $ m $.
This result will be used to derive an upper bound on the cardinality of optimal multiset
codes with specified minimum distance $ d $ and alphabet size $ q = m + 1 $
(Theorem \ref{thm:bounds} in Section \ref{sec:bounds}).
\begin{IEEEproof}
  The bound in \eqref{eq:mu_upper} follows from Theorem \ref{thm:codesaspackings} and
is a version of the code-anticode bound \cite{delsarte, ahlswede} adapted to the space
studied here.
Namely, if $ \da(\C) = d > r^\sml{+} + r^\sml{-} $, then every translate of
$ S_{m}(r^\sml{+}, r^\sml{-}) $ in $ \mathbb{Z}^{m} $ contains at most one codeword from
$ \C $, and hence $ \overline{\mu}(\C) \cdot | S_{m}(r^\sml{+}, r^\sml{-}) | \leq 1 $.

By Lemma \ref{thm:Anball} we then have, for $ m \geq r^\sml{+} $,
\begin{equation}
\begin{aligned}
 \overline{\mu}(\C)^{-1}
    &\geq \sum_{ j = 0 }^{ r^\sml{+} }   \binom{m}{j}
                                         \binom{r^\sml{+}}{j}
                                         \binom{r^\sml{-} + m - j}{m - j}  \\
    &\geq \sum_{ j = 0 }^{ r^\sml{+} } \frac{ (m - j + 1)^j }{ j! } \binom{r^\sml{+}}{j} \frac{ (m - j + 1)^{r^\sml{-}} }{ r^\sml{-}! }  \\
    &> \frac{ (m - r^\sml{+} + 1)^{r^\sml{+} + r^\sml{-}} }{ r^\sml{+}! \ r^\sml{-}! } ,
\end{aligned}
\end{equation}
where we used $ \binom{m}{j} \geq \frac{(m-j+1)^j}{j!} $, and the last inequality
is obtained by keeping only the summand $ j = r^\sml{+} $.
Taking $ r^\sml{+} = \ceil{\frac{d-1}{2}} $ and $ r^\sml{-} = \floor{\frac{d-1}{2}} $,
we get \eqref{eq:phik}.

Similarly, for $ 0 \leq r^\sml{+} - m \leq r^\sml{-} $, we have
\begin{equation}
\begin{aligned}
 \overline{\mu}(\C)^{-1}
    &\geq \sum_{ j = 0 }^{ m }  \binom{m}{j}
                                \binom{r^\sml{+}}{j}
                                \binom{r^\sml{-} + m - j}{m - j}  \\
    &\geq \sum_{ j = 0 }^{ m } \binom{m}{j} \frac{ (r^\sml{+} - j + 1)^j }{ j! } \frac{ (r^\sml{-} + 1)^{(m-j)} }{ (m-j)! }  \\
    &=    \frac{1}{m!} \sum_{ j = 0 }^{ m } \binom{m}{j}^2 (r^\sml{+} - j + 1)^j (r^\sml{-} + 1)^{(m-j)}  \\
    &> \frac{ (r^\sml{+} - m + 1)^m }{ m! } \binom{2m}{m} .
\end{aligned}
\end{equation}
In the last step we used the assumption $ r^\sml{-} \geq r^\sml{+} - m $ and
the identity $ \sum_{j=0}^{m} \binom{m}{j}^2 = \binom{2m}{m} $.
Letting $ r^\sml{+} = \big\lfloor \frac{d - 1 + m}{2} \big\rfloor $ and
$ r^\sml{+} + r^\sml{-} = d - 1 $, we get \eqref{eq:phih}.
\end{IEEEproof}

\subsection{Codes in $ (\mathbb{Z}^{m}, \da) $: Construction based on $ B_h $ sets}
\label{sec:AqBh}

In this subsection, we describe a method of construction of codes in $ (\mathbb{Z}^m, \da) $
having a given minimum distance.
As we shall demonstrate in Section \ref{sec:perfect}, the construction is optimal
for some sets of parameters, and in fact produces perfect or diameter-perfect codes
in those instances.

Let $ G $ be an Abelian group of order $ v $, written additively.
A set $ B = \{ b_0, b_1, \ldots, b_{m} \} \subseteq G $ is said to be a
\emph{$ B_h $ set} (or \emph{$ B_h $ sequence}, or \emph{Sidon set} of order $ h $)
if the sums $ b_{i_1} + \cdots + b_{i_h} $,
$ 0 \leq i_1 \leq \cdots \leq i_h \leq m $, are all different.
If $ B $ is a $ B_h $ set, then so is $ B - b_0 \equiv \{ 0, b_1-b_0, \ldots, b_m-b_0 \} $,
and vice versa;
we shall therefore assume in the sequel that $ b_0 = 0 $.
With this convention, the requirement for $ B $ to be a $ B_h $ set is that
the sums $ b_{i_1} + \cdots + b_{i_u} $ are different for all $ u \in \{0, 1, \ldots, h\} $
and $ 1 \leq i_1 \leq \cdots \leq i_u \leq m $.
Among the early works on these and related objects we mention
Singer's construction \cite{singer} of optimal $ B_2 $ sets in
$ \mathbb{Z}_v \defeq \mathbb{Z} / v\mathbb{Z} $, for $ m $ a prime power and
$ v = m^2 + m + 1$, and a construction by Bose and Chowla \cite{bose+chowla}
of $ B_h $ sets in $ \mathbb{Z}_v $ for arbitrary $ h \geq 1 $ when:
\begin{inparaenum}[1)]
\item  $ m $ is a prime power and $ v = m^h + m^{h-1} + \cdots + 1 $, and
\item  $ m + 1 $ is a prime power and $ v = (m+1)^h - 1 $.
\end{inparaenum}
Since these pioneering papers, research in the area has become quite extensive,
see \cite{obryant} for references,
and has also found various applications in coding theory, e.g.,
\cite{barg, derksen, graham+sloane, klove, levenshtein, varshamov}.

The following theorem states that linear codes in $ (\mathbb{Z}^m, \da) $
(or, equivalently, in $ (A_m, d_1) $) are in fact geometric analogs of $ B_h $ sets.

\begin{theorem}
\label{thm:Bh}
  Let $ h \geq 1$ be an integer.
\begin{itemize}
\item[(a)]
Assume that $ B = \{ 0, b_1, \ldots, b_m \} $ is a $ B_h $ set in an Abelian
group $ G $ of order $ v $, and that $ B $ generates $ G $.
Then
\begin{equation}
\label{eq:Bhlattice}
  \Lat = \left\{ {\bf x} \in \mathbb{Z}^m : \sum_{i=1}^{m} x_i b_i = 0 \right\}
\end{equation}
is a linear code of minimum distance $ \da(\Lat) > h $ and density $ \mu(\Lat) = \frac{1}{v} $
in $ \mathbb{Z}^m $.
(Here $ x_i b_i $ denotes the sum in $ G $ of $ |x_i| $ copies of $ b_i $ if $ x_i > 0 $,
or $ -b_i $ if $ x_i < 0 $.)
\item[(b)]
Conversely, if $ \Lat' \subseteq \mathbb{Z}^m $ is a linear code of minimum distance
$ \da(\Lat') > h $, then the group $ G = \mathbb{Z}^m / \Lat' $ contains a $ B_h $ set
of cardinality $ m + 1 $ that generates $ G $.
\end{itemize}
\end{theorem}
\begin{IEEEproof}
  Both statements follow from Theorem \ref{thm:codesaspackings} and the familiar group-theoretic
formulation of lattice packing problems \cite{stein67, hamaker, stein74, galovich, hickerson, stein84, stein+szabo},
so we only sketch the proof of {\it (a)}.

If $ B = \{ 0, b_1, \ldots, b_m \} $ is a $ B_h $ set, then $ (S_m(h, 0), \Lat) $ is
a packing in $ \mathbb{Z}^m $, or, in coding-theoretic terminology, the ``error-vectors''
from $ S_m(h, 0) $ are correctable and have different syndromes.
Namely, we see from \eqref{eq:Bhlattice} that the syndromes are of the form
$ b_{i_1} + \cdots + b_{i_u} $, where $ u \in \{0, 1, \ldots, h\} $ and
$ 1 \leq i_1 \leq \cdots \leq i_u \leq m $, and the condition that they are all different
is identical to the condition that $ \{ 0, b_1, \ldots, b_m \} $ is a $ B_h $ set.
This implies $ \da(\Lat) > h $ (see Theorem \ref{thm:codesaspackings}).
Furthermore, if $ B $ generates $ G $, then $ G $ is isomorphic to $ \mathbb{Z}^m / \Lat $,
meaning that $ \mu(\Lat) = \frac{1}{|G|} $.
\end{IEEEproof}

\begin{figure}[h]
 \centering
  \includegraphics[width=\columnwidth]{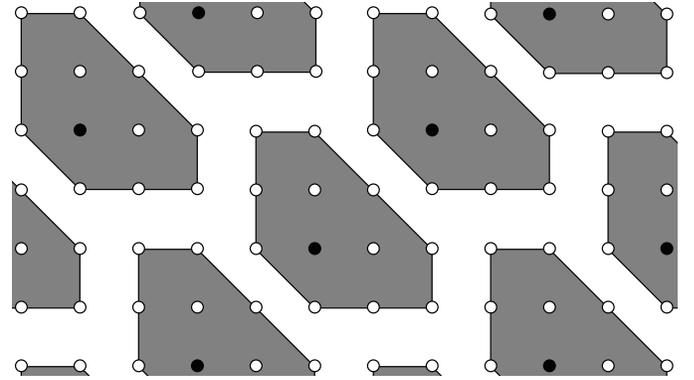}
\caption{Tiling of $ \mathbb{Z}^2 $ by the anticode $ S_2(2,1) $.}
\label{fig:S2_tiling}
\end{figure}

\begin{example}
\label{ex:Bh}
\textnormal{
  The set $ \{ (0,0), (1,1), (0,5) \} $ is a $ B_3 $ set in the group $ \mathbb{Z}_2 \times \mathbb{Z}_6 $.
The corresponding lattice packing $ (S_2(2,1), \Lat) $ in $ \mathbb{Z}^2 $ is
illustrated in Figure \ref{fig:S2_tiling}.
Notice that this is in fact a perfect packing, i.e., a tiling of the grid $ \mathbb{Z}^2 $.
This means that $ \Lat $ is a diameter-perfect linear code of minimum distance
$ \da(\Lat) = 4 $; see Section \ref{sec:diameter} ahead.
}
\myqed
\end{example}

The significance of Theorem \ref{thm:Bh} is twofold.
First, using the known constructions of $ B_h $ sets one automatically obtains
good codes in $ (\mathbb{Z}^m, \da) $.
This in particular gives a lower bound on the achievable density of codes in
$ (\mathbb{Z}^m, \da) $, i.e., the ``achievability'' counterpart of Theorem
\ref{thm:boundsZq}.
For example, a construction of Bose and Chowla mentioned above asserts the
existence of linear codes in $ (\mathbb{Z}^m, \da) $ of minimum distance $ d $
and density $ > (m+1)^{1-d} $, for any $ d \geq 2 $ and $ m \geq 1 $ with $ m + 1 $
a prime power.
Second, this geometric interpretation enables one to derive the best known bounds
on the parameters of $ B_h $ sets.
Namely, having in mind the correspondence between the density of the code and the
size of the group containing a $ B_h $ set, between the minimum distance of the
code and the parameter $ h $, and between the dimension of the code and the
cardinality of the $ B_h $ set, one can restate the inequalities from
Theorem \ref{thm:boundsZq} in terms of the parameters of $ B_h $ sets.
The resulting bounds are either equivalent to, or improve upon the known bounds%
\footnote{To the best of our knowledge, the best known bounds for $ B_h $ sets
in finite groups are stated in \cite{jia, chen}.}:
The bound in \eqref{eq:phik} is equivalent to those in \cite[Thm 2]{jia} and
\cite[Thm 2]{chen}, but with an explicit error term, while the bound in
\eqref{eq:phih} improves upon that in \cite[Thm 1(v)]{jia} by a factor of two.
See \cite{kovacevic+tan} for an explicit statement of these bounds and their
further improvements.

\subsection{Perfect codes in $ (\mathbb{Z}^m, \da) $}
\label{sec:perfect}

A code is said to be $ r $-\emph{perfect} if balls of radius $ r $ around the
codewords are disjoint and cover the entire space.
Perfect codes are the best possible codes having a given error correction radius;
it is therefore important to study their existence, and methods of construction
when they do exist.
Notice that, by Theorem \ref{thm:Bh}, linear $ r $-perfect codes in $ (\mathbb{Z}^m, \da) $
correspond to $ B_{2r} $ sets of cardinality $ m + 1 $ in Abelian groups of order
$ v = |S_m(r,r)| $.

\vspace{2mm}

\subsubsection{$ {1} $-Perfect codes and planar difference sets}
\label{sec:planar}

Linear $ 1 $-perfect codes in $ (\mathbb{Z}^m, \da) $ correspond to $ B_2 $ sets
of cardinality $ m + 1 $ in Abelian groups of order $ v = |S_m(1,1)| = m^2 + m + 1 $.
Such sets are better known in the literature as \emph{planar} (or \emph{simple})
\emph{difference sets}.
The condition that all the sums $ b_i + b_j $ are different, up to the order of
the summands, is equivalent to the condition that all the differences $ b_i - b_j $,
$ i \neq j $, are different (hence the name), and the condition that the order of
the group is $ v = m^2 + m + 1 $ means that \emph{every} nonzero element of the
group can be expressed as such a difference.		
If $ D $ is a planar difference set of size $ m + 1 $, then $ m $ is referred to
as the \emph{order} of $ D $.
If $ G $ is Abelian, cyclic, etc., then $ D $ is also said to be Abelian, cyclic,
etc., respectively.

Planar difference sets and their generalizations are extensively studied objects
\cite{designs}, and have also been applied in communications and coding theory in
various settings; see for example \cite{atkinson, lam+sarwate, macwilliams+sloane, ding}.
The following claim, which is a corollary to Theorem \ref{thm:Bh}, states that these
objects are essentially equivalent to linear $ 1 $-perfect codes in $ (\mathbb{Z}^m, \da) $,
and can be used to construct the latter via \eqref{eq:Bhlattice}.

\begin{theorem}
\label{thm:1perfect}
  The space $ (\mathbb{Z}^m, \da) $ admits a linear $ 1 $-perfect code if and only
if there exists an Abelian planar difference set of order $ m $.
\hfill \IEEEQED
\end{theorem}

\begin{figure}[h]
 \centering
  \subfigure[The code viewed in the plane $ x_0 = 0 $.]
  {
   \includegraphics[width=.99\columnwidth]{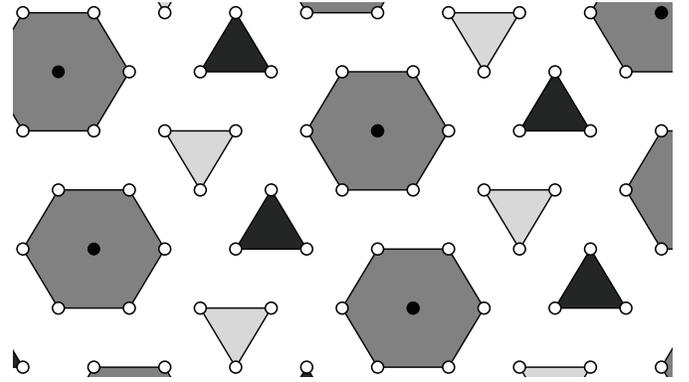}
   \label{fig:A3perfect}
  }
	\vspace{1mm}
  \subfigure[Intersections of a ball in $ (A_3, d_1) $ with the planes $ x_0 = \text{const} $.]
  {
	 \makebox[\columnwidth]{
    \includegraphics[width=0.35\columnwidth]{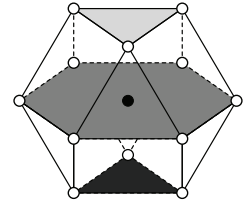}
    \label{fig:ball3color}
	 }
	}
\caption{$ 1 $-perfect code in $ (A_3, d_1) $.}
\end{figure}

\begin{example}
\textnormal{
  Consider a planar difference set $ \{0, 1, 3, 9\} $ in the cyclic group $ \mathbb{Z}_{13} $.
The corresponding $ 1 $-perfect code in $ (A_3, d_1) $ is illustrated in Figure \ref{fig:A3perfect}.
The figure shows the intersection of $ A_3 $ with the plane $ x_0 = 0 $.
Intersections of a ball of radius $ 1 $ in $ (A_3, d_1) $---a cuboctahedron---with the
planes $ x_0 = \text{const} $ are shown in Figure \ref{fig:ball3color} for clarification.
}
\myqed
\end{example}

Abelian planar difference sets of prime power orders $ m $ were first constructed by
Singer \cite{singer}.
It is believed that this condition on $ m $ is in fact necessary, but this question---%
now known as the prime power conjecture \cite[Conj. 7.5, p. 346]{designs}---remains
open for nearly eight decades.
By Theorem \ref{thm:1perfect}, the statement can be rephrased as follows:

\begin{conjecture}[{Prime power conjecture}]
\label{conj:ppc}
  There exists a linear $ 1 $-perfect code in $ (\mathbb{Z}^m, \da) $ if and only if
the dimension $ m $ is a prime power.
\myqed
\end{conjecture}

\subsubsection{$ {r} $-Perfect codes in $ {(\mathbb{Z}^m, \da)} $}
\label{sec:rperfect}

Theorem \ref{thm:1perfect}, together with a direct inspection of the one- and
two-dimensional cases (see also \cite{costa}), yields the following fact:

\begin{theorem}
\label{thm:perfect}
  There exists an $ r $-perfect code in $ (\mathbb{Z}^m, \da) $ for:%
\begin{itemize}
\item  $ m \in \{1, 2\} $, $ r $ arbitrary;
\item  $ m \geq 3 $ a prime power, $ r = 1 $.   \hfill \IEEEQED
\end{itemize}
\end{theorem}

Proving (non-)existence of perfect codes for arbitrary pairs $ (m, r) $ seems to be a
highly non-trivial problem%
\footnote{It should also be contrasted with the well-known Golomb--Welch conjecture
\cite{golomb+welch} (see also, e.g., \cite{horak}) stating that $ r $-perfect
codes in $ \mathbb{Z}^m $ under the $ \ell_1 $ metric exist only in the following cases:
\begin{inparaenum}[1)]
\item  $ m \in \{1, 2\} $, $ r $ arbitrary, and 
\item  $ r = 1 $, $ m $ arbitrary.
\end{inparaenum}
}.
We shall not be able to solve it here, but Theorem \ref{thm:rperfect} below is a
step in this direction.

\begin{figure}%[h]
 \centering
  \includegraphics[width=0.75\columnwidth]{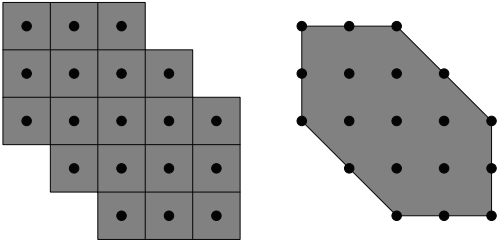}
\caption{Bodies in $ \mathbb{R}^2 $ corresponding to a ball of radius $ 2 $
         in $ \left( \mathbb{Z}^2, \da \right) $:
         The cubical cluster $ S^{\cube}_2(2) $ (left) and the convex hull $ S^{\conv}_2(2) $ (right).}
\label{fig:Z2tile}
\end{figure}

For $ S \subset \mathbb{Z}^m $, denote by $ S^{\cube} $ the body in
$ \mathbb{R}^m $ defined as the union of unit cubes translated to the points
of $ S $, namely, $ S^{\cube} = \bigcup_{{\bf y} \in S} ( {\bf y} + [-1/2, 1/2]^m ) $,
and by $ S^{\conv} $ the convex hull in $ \mathbb{R}^m $ of the points in
$ S $ (see Figure \ref{fig:Z2tile}).

\begin{lemma}
\label{thm:volumes}
  Let $ S_m(r) \equiv S_m(r,r) $ be the ball of radius $ r $ around $ \bf 0 $
in $ ( \mathbb{Z}^m, \da ) $.
The volumes of the bodies $ S^{\cube}_m(r) $ and $ S^{\conv}_m(r) $
are given by
\begin{align}
\label{eq:Dn}
   \vol(S^{\cube}_m(r)) &= \sum_{ j = 0 }^{ \min\{ m, r \} }  \binom{m}{j}
                                                              \binom{r}{j}
                                                              \binom{r + m - j}{m - j} \\ %\quad , \quad
\label{eq:volconv}
   \vol(S^{\conv}_m(r)) &= \frac{r^m}{m!} \binom{2m}{m} .
\end{align}
Furthermore, $ \lim_{r \to \infty} \vol\left(S^{\conv}_m(r)\right) / \vol\left(S^{\cube}_m(r)\right) = 1 $
for any fixed $ m \geq 1 $.
\end{lemma}
\begin{IEEEproof}
  $ S^{\cube}_m(r) $ consists of $ |S_m(r)| $ unit cubes so \eqref{eq:Dn}
follows from Lemma \ref{thm:Anball}.

To compute the volume of $ S^{\conv}_m(r) $, observe its intersection with the orthant
$ x_{1}, \ldots, x_j > 0 $, $ x_{j+1}, \ldots, x_{m} \leq 0 $, where $ 0 \leq j \leq m $.
The volume of this intersection is the product of the volumes of the $ j $-simplex
$ \big\{ (x_1, \ldots, x_j) : x_i > 0, \linebreak \sum_{i=1}^j x_i \leq r \big\} $, which
is known to be $ r^j/j! $, and of the $ (m-j) $-simplex
$ \big\{ (x_{j+1}, \ldots, x_m) : x_i \leq 0, \sum_{i=j+1}^m x_i \geq -r \big\} $, which
is $ r^{m-j}/(m-j)! $.
This implies that
$ \vol\left(S^{\conv}_m(r)\right) = \sum_{j=0}^m \binom{m}{j} \frac{r^j}{j!} \frac{r^{m-j}}{(m-j)!} $,
which reduces to \eqref{eq:volconv} by using the identity
$ \sum_{j=0}^m \binom{m}{j}^2 = \binom{2m}{m} $.

Finally, when $ r \to \infty $ we have $ \binom{r}{j} \sim \frac{r^j}{j!} $ and so
$ \vol\left(S^{\cube}_m(r)\right)  \sim  \frac{r^m}{m!} \binom{2m}{m}  =  \vol\left(S^{\conv}_m(r)\right) $.
(Here $ f(r) \sim g(r) $ stands for $ \lim_{r \to \infty} f(r) / g(r) = 1 $.)
\end{IEEEproof}

\begin{theorem}
\label{thm:rperfect}
  There are no $ r $-perfect codes in $ ( \mathbb{Z}^m, \da ) $,\linebreak $ m \geq 3 $,
for large enough $ r $, i.e., for $ r \geq r_0(m) $.
\end{theorem}
\begin{IEEEproof}
 The proof relies on the idea used to prove the corresponding statement for $ r $-perfect
codes in $ \mathbb{Z}^m $ under $ \ell_1 $ distance \cite{golomb+welch}.
First observe that an $ r $-perfect code in $ ( \mathbb{Z}^m, \da ) $ would
induce a tiling of $ \mathbb{R}^m $ by $ S^{\cube}_m(r) $, and a \emph{packing}
by $ S^{\conv}_m(r) $.
The relative efficiency of the latter with respect to the former is defined as
the ratio of the volumes of these bodies, $ \vol\left(S^{\conv}_m(r)\right) / \vol\left(S^{\cube}_m(r)\right) $,
which by Lemma \ref{thm:volumes} converges to $ 1 $ as $ r $ tends to infinity.
This has the following consequence: If an $ r $-perfect code exists in
$ (\mathbb{Z}^m, \da) $ for infinitely many $ r $, then there exists a
tiling of $ \mathbb{R}^m $ by translates of $ S^{\cube}_m(r) $ for infinitely
many $ r $, which further implies that a packing of $ \mathbb{R}^m $ by translates of
$ S^{\conv}_m(r) $ exists which has efficiency arbitrarily close to $ 1 $.
But then there would also be a packing by $ S^{\conv}_m(r) $ of efficiency $ 1 $,
i.e., a tiling (in \cite[Appendix]{golomb+welch} it is shown that there exists a
packing whose density is the supremum of the densities of all possible packings with
a given body).
This is a contradiction.
Namely, it is known \cite[Thm 1]{mcmullen} that a necessary condition for a convex
body to be able to tile space is that it be a polytope with centrally symmetric%
\footnote{A polytope $ P \subset \mathbb{R}^m $ is centrally symmetric if
its translation $ \tilde{P} = P - {\bf c} $ satisfies $ \tilde{P} = -\tilde{P} $
for some $ {\bf c} \in \mathbb{R}^m $.}
facets, which $ S^{\conv}_m(r) $ fails to satisfy for $ m \geq 3 $.
For example, the facet which is the intersection of $ S^{\conv}_m(r) $ with
the hyperplane $ x_1 = -r $ is the simplex
$ \big\{ (x_2, \ldots, x_m) : x_i \geq 0,\, \sum_{i=2}^m x_i \leq r \big\} $,
a non-centrally-symmetric body.
\end{IEEEproof}

\vspace{2mm}

\subsubsection{Diameter-perfect codes in $ {(\mathbb{Z}^m, \da)} $}
\label{sec:diameter}

The following generalization of a notion of perfect code, adjusted to our setting,
was introduced in \cite{ahlswede}.
We say that a code $ \C \subseteq \mathbb{Z}^m $ of minimum distance $ \da(\C) $
is \emph{diameter-perfect} if there exists an anticode $ S \subset \mathbb{Z}^m $
of diameter $ \da(\C) - 1 $ such that $ \mu(\C) \cdot |S| = 1 $.
Namely, by the arguments from \cite{delsarte, ahlswede}, we know that for any such
code-anticode pair, we must have $ \mu(\C) \cdot |S| \leq 1 $ (see also Theorems
\ref{thm:codesaspackings} and \ref{thm:boundsZq}).
Therefore, a code is said to be diameter-perfect if it achieves this bound.
This notion is especially interesting when the minimum distance of a code is even,
which can never be the case for perfect codes.

\begin{theorem}
\label{thm:diamperfect}
  There exists a diameter-perfect code of minimum distance $ 2 r $ in $ (\mathbb{Z}^m, \da) $ for:
\begin{itemize}
\item  $ m \in \{1, 2\} $, $ r $ arbitrary;
\item  $ m \geq 3 $, $ r = 1 $.
\end{itemize}
\end{theorem}
\begin{IEEEproof}
We show that, in all the stated cases, there exists a lattice tiling
$ (S_m(r,r-1), \Lat) $, $ \Lat \subseteq \mathbb{Z}^m $.
This will prove the claim because $ S_m(r,r-1) $ is an anticode of diameter
$ 2 r - 1 $.
Dimension $ m = 1 $ is trivial.
In dimension $ m = 2 $, one can check directly that the lattice generated by
the vectors $ (r, r) $ and $ (0, 3r) $ defines a tiling by $ S_2(r,r-1) $ for
any $ r \geq 1 $, see Figure \ref{fig:S2_tiling}.
It can also be shown that this lattice is unique---there are no other diameter-perfect
codes of minimum distance $ 2 r $ in $ (\mathbb{Z}^2, \da) $.
The statement for $ r = 1 $ is left.
By Theorem \ref{thm:Bh}, a lattice tiling $ (S_m(1,0), \Lat) $,
$ \Lat \subseteq \mathbb{Z}^m $, exists if and only if a $ B_{1} $ set exists
in some Abelian group $ G $ of order $ |S_m(1,0)| = m + 1 $.
Notice that any $ G $ is itself such a set.
\end{IEEEproof}

\vspace{2mm}

In dimensions $ m \geq 3 $, one can show in a way analogous to Theorem \ref{thm:rperfect}
that tilings of $ \mathbb{Z}^m $ by the anticodes $ S_m(r,r-1) $ do not exist for
$ r $ large enough.

\section{Multiset codes: Construction and bounds}
\label{sec:bounds}

In this section we describe a construction of multiset codes over an arbitrary
$ q $-ary alphabet, i.e., codes in the simplex $ \triangle_n^{q-1} $.
We then derive bounds on the cardinalities of optimal multiset codes and examine their
asymptotic behavior in several regimes of interest.
These bounds will, in particular, demonstrate optimality of the presented construction
for some sets of parameters.

\subsection{Construction based on Sidon sets}
\label{sec:construction}

The construction given next is inspired by the observation in Theorem \ref{thm:Bh},
which states that linear codes in $ A_{q-1} $ lattices are essentially equivalent to
Sidon sets in Abelian groups.

Let $ B = \{ b_0, b_1, \ldots, b_{q-1} \} \subseteq G $ and $ b \in G $, where $ G $ is
an Abelian group.
Define
\begin{equation}
\label{eq:Bhcode}
  \C_{n}^{\sml{(G, B, b)}} = \left\{ {\bf x} \in \triangle^{q-1}_n  :  \sum_{i=0}^{q-1} x_i b_i = b \right\} .
\end{equation}

\begin{theorem}
\label{thm:Bhcodes}
  If $ B $ is a $ B_h $ set, then the code $ \C_{n}^{\sml{(G, B, b)}} $ can
correct $ h $ deletions.
\end{theorem}

In other words, if $ B $ is a $ B_h $ set and $ \big| \C_{n}^{\sml{(G, B, b)}} \big| \geq 2 $,
then $ d\big(\C_{n}^{\sml{(G, B, b)}}\big) > h $ (see Theorem \ref{thm:corrdel}).
\vspace{2mm}

\begin{IEEEproof}
  Suppose that $ \C_{n}^{\sml{(G, B, b)}} $ cannot correct $ h $ deletions, i.e.,
there exist two different codewords $ {\bf x}, {\bf y} $ and two different vectors
$ {\bf f}, {\bf g} $ such that $ f_i, g_i \geq 0 $, $ \sum_{i=0}^{q-1} f_i = \sum_{i=0}^{q-1} g_i = h $,
and $ {\bf x} - {\bf f} = {\bf y} - {\bf g} $.
This implies that $ \sum_{i=0}^{q-1} (x_i - f_i) b_i =\linebreak \sum_{i=0}^{q-1} (y_i - g_i) b_i $
and, since $ \sum_{i=0}^{q-1} x_i b_i = \sum_{i=0}^{q-1} y_i b_i = b $, we get
$ \sum_{i=0}^{q-1} f_i b_i = \sum_{i=0}^{q-1} g_i b_i $.
This means that the set $ \{ b_0, b_1, \ldots, b_{q-1} \} $ is not a $ B_h $ set.
\end{IEEEproof}

\begin{theorem}
\label{thm:linear}
  Let $ \C = (\Lat + {\bf t}) \cap \triangle^{q-1}_n $ be a linear multiset code
of minimum distance $ d_1(\C) > h $, where $ \bf t $ satisfies $ t_i \geq h $ for
all $ i \in \alf{q} $ (and hence $ n \geq h q $).
Then $ \C $ is necessarily of the form \eqref{eq:Bhcode} for some $ B_{h} $
set $ \{b_0, b_1, \ldots, b_{q-1}\} $.
\end{theorem}
\begin{IEEEproof}
  Let $ \C = (\Lat + {\bf t}) \cap \triangle^{q-1}_n $ be a linear code of
minimum distance $ d_1(\C) > h $.
Notice that $ {\bf t} \in \C $ since $ {\bf 0} \in \Lat $.\linebreak
If $ \bf t $ satisfies the condition $ t_i \geq h $ then the entire ball of\linebreak radius $ h $
around $ \bf t $ (regarded in $ A_{q-1} + {\bf t} $) belongs to $ \triangle^{q-1}_n $,
i.e., all the points in this ball have non-negative coordinates.
Moreover, since the code $ \C $ has distance $ > h $, this ball does not
contain another codeword of $ \C $.
These two facts imply that $ \Lat $ is a linear code of minimum distance
$ d_1(\Lat) > h $ in $ A_{q-1} $.
The claim then follows by invoking Theorem \ref{thm:Bh} which states that any
such code is of the form $ \big\{ {\bf x} \in A_{q-1} : \sum_{i=0}^{q-1} x_i b_i = 0 \big\} $,
where $ \{b_0, b_1, \ldots, b_{q-1}\} $ is a $ B_{h} $ set in the quotient group $ A_{q-1} / \Lat $.
\end{IEEEproof}

\begin{remark}
\label{rem:linear}
\textnormal{
  Suppose $ \{ \C'_n \}_n $ is a family of linear multiset codes obtained from a
family of lattices $ \{ \Lat'_n \}_n $, $ \Lat'_n \subseteq A_{q-1} $, where
$ n $ denotes the code block-length.
If the density of $ \Lat'_n $ is bounded, meaning that $ \mu(\Lat'_n) = {\mathcal O}(1) $
as $ n \to \infty $, then every code $ \C'_n $ of sufficiently large block-length ($ n \geq n_0 $)
will necessarily contain a codeword $ \bf t $ satisfying $ t_i \geq d_1(\C'_n) - 1 $, and
will by Theorem \ref{thm:linear} be of the form \eqref{eq:Bhcode}.
}
\myqed
\end{remark}

\subsection{Bounds and asymptotics: Fixed alphabet case}
\label{sec:boundsqfixed}

Let $ M_{q}(n, h) $ denote the cardinality of the largest multiset code of
block-length $ n $ over the alphabet $ \alf{q} $ which can correct $ h $ deletions
(or, equivalently, which has minimum distance $ > h $), and $ M^\textsc{l}_{q}(n,h) $
the cardinality of the largest \emph{linear} multiset code with the same parameters.
We shall assume in the following that $ n > h $; this condition is necessary and
sufficient for the existence of nontrivial codes of distance $ > h $, i.e., codes
with at least two codewords.

When discussing asymptotics, the following conventions will be used:
$ f(n) \sim g(n) $ means $ \lim_{n \to \infty} f(n) / g(n) = 1 $,
and $ f(n) \gtrsim g(n) $ means $ \liminf_{n \to \infty} f(n) / g(n) \geq 1 $.

\pagebreak
For notational convenience, denote by $ \beta(h, q) $ the size of the anticode
$ S_{q-1}\left(\lceil \frac{h}{2} \rceil, \lfloor \frac{h}{2} \rfloor\right) \subset \mathbb{Z}^{q-1} $
of diameter $ h $ (see \eqref{eq:Anball}).
That is,
\begin{equation}
\label{eq:Anball2}
  \beta(h, q) =
     \sum_{ j \geq 0 }
			  \binom{q-1}{j}
        \binom{\lceil  \frac{h}{2} \rceil}{j}
        \binom{\lfloor \frac{h}{2} \rfloor + q-1 - j}{q-1 - j} .
\end{equation}
Let also $ \phi(h,q) $ denote the order of the smallest Abelian group containing
a $ B_{h} $ set of cardinality $ q $.
The lower bounds that follow will be expressed in terms of this quantity; more
explicit lower bounds stated in terms of the parameters $ h, q $ can be obtained
from the known upper bounds on $ \phi(h,q) $ \cite{singer, bose+chowla, jia, kovacevic+tan}.

\begin{theorem}
\label{thm:bounds}
For every $ q \geq 2 $ and $ n > h \geq 1 $,
\begin{align}
\label{eq:lowerbound}
  &M_{q}(n,h)  \geq  \frac{ \binom{n + q - 1}{q - 1} }{ \phi(h, q) } ,  \\
\label{eq:upperboundq}
	&M_{q}(n,h)  \leq  \frac{ \binom{n + q - 1}{q - 1} }{ \beta(h, q) }  + 
	                   \sum_{j=1}^{q \lceil \frac{h}{2} \rceil} \binom{n + q-1 - j}{q - 2}  .
\end{align}
\end{theorem}
\begin{IEEEproof}
  The lower bound in \eqref{eq:lowerbound} follows from the construction described
in the previous subsection.
Fix $ n $, an Abelian group $ G $, and a $ B_h $ set $ B \subseteq G $ with $ |B| = q $.
Then by Theorem \ref{thm:Bhcodes} the codes $ \C_{n}^{\sml{(G, B, b)}} $
can correct $ h $ deletions.
Furthermore, since they form a partition of $ \triangle^{q-1}_n $, and since there
are $ |G| $ of them (one for each $ b \in G $), at least one has cardinality
$ \geq |\triangle^{q-1}_n| / |G| $.
To get the tightest bound take $ G $ to be the smallest group containing a $ B_h $
set with $ q $ elements, i.e., $ |G| = \phi(h,q) $.

The upper bound in \eqref{eq:upperboundq} follows from an argument similar to the one
that led to the code-anticode bound \eqref{eq:mu_upper} in Theorem \ref{thm:boundsZq}.
The difficulty in directly applying that argument to codes in $ \triangle^{q-1}_n $
is that, if a codeword $ \bf x $ is too close to the ``boundary'' of $ \triangle^{q-1}_n $,
then the corresponding anticode around $ {\bf x} $ will be ``clipped'' (see Figure
\ref{fig:simplex}) and will have cardinality smaller than $ \beta(h,q) $.
The vectors $ {\bf x} \in \triangle^{q-1}_n $ for which this is not true, i.e., anticodes
around which have cardinality $ \beta(h,q) $, are those satisfying $ x_i \geq \lceil \frac{h}{2} \rceil $
for all $ i \in \alf{q} $.
The set of such sequences can be written as
$ \left(\lceil \frac{h}{2} \rceil, \ldots, \lceil \frac{h}{2} \rceil\right) + \triangle^{q-1}_{n'} $,
where $ n' = n - q \lceil \frac{h}{2} \rceil $, and is of cardinality $ | \triangle^{q-1}_{n'} | $.
Now, write $ M_{q}(n,h) = M' + M'' $, where $ M' $ is the number of codewords of
an optimal code that belong to
$ \left(\lceil \frac{h}{2} \rceil, \ldots, \lceil \frac{h}{2} \rceil\right) + \triangle^{q-1}_{n'} $,
and $ M'' $ the number of remaining codewords.
By the code-anticode argument leading to \eqref{eq:mu_upper}, we have
$ M' \cdot \beta(h,q) \leq |\triangle^{q-1}_n| $,
which gives the first summand in the upper bound \eqref{eq:upperboundq}.
The second summand is the size of the remaining part of the simplex,
$ |\triangle^{q-1}_n| - |\triangle^{q-1}_{n'}| =
  \sum_{j=1}^{n-n'} \binom{n + q-1 - j}{q - 2} $,
which is certainly an upper bound on $ M'' $.
\end{IEEEproof}

\vspace{2mm}

In the asymptotic case, as the block-length grows to infinity, we get the following bounds.

\begin{theorem}
\label{thm:asympboundsq}
  For every fixed $ q \geq 2 $ and $ h \geq 1 $, as $ n \to \infty $,
\begin{align}
\label{eq:asympq}
  \frac{ n^{q-1} }{ (q-1)!\ \beta(h, q) }  \gtrsim\  &M_{q}(n,h)  \gtrsim  \frac{ n^{q-1} }{ (q-1)!\ \phi(h, q) }  \\
\label{eq:asympql}
	&M^\textsc{l}_{q}(n,h)  \sim  \frac{ n^{q-1} }{ (q-1)!\ \phi(h, q) }  .
\end{align}
\end{theorem}
\begin{IEEEproof}
  \eqref{eq:asympq} follows from \eqref{eq:lowerbound} and \eqref{eq:upperboundq}
after observing that $ \binom{n + q - 1}{q - 1}  \sim  \frac{ n^{q-1} }{ ({q-1})! } $ and
that the second summand in \eqref{eq:upperboundq} is of the order $ {\mathcal  O}(n^{q-2}) $.

Theorem \ref{thm:Bh} implies that $ 1 / \phi(h, q) $ is the largest density
a sublattice of $ A_{q-1} $ with minimum distance $ > h $ can have.
Since the dimension $ q-1 $ and the minimum distance $ h + 1 $ are fixed, and the
size of the simplex grows indefinitely (as $ n \to \infty $), it follows that
$ M^\textsc{l}_{q}(n,h)  \sim  |\triangle^{q-1}_n| / \phi(h, q) $.
This proves \eqref{eq:asympql}.
\end{IEEEproof}

\vspace{2mm}

We see from Theorem \ref{thm:asympboundsq} that the cardinality of optimal multiset
codes over a fixed alphabet scales as $ \Theta(n^{q-1}) $.
Therefore, at most polynomially many codewords are available to the transmitter in
this setting, as opposed to exponentially many codewords available in the standard
models where the transmitter sends \emph{sequences} of symbols.
This is the price paid for storing/transmitting information in an unordered way.
Our main contribution in Theorem \ref{thm:asympboundsq} is to establish bounds on
the implied constant in the $ \Theta(n^{q-1}) $ term for general multiset codes
and the exact implied constant in the same term for linear multiset codes.
These constants of course depend on the number of deletions $ h $ and the alphabet
size $ q $.

The following claim states that the construction described in Section \ref{sec:construction}
produces asymptotically optimal multiset codes over binary and ternary alphabets
for arbitrary minimum distance, and over arbitrary alphabets for small distances.
These codes are in fact \emph{asymptotically (diameter-)perfect}, meaning that they
asymptotically achieve the upper bound \eqref{eq:upperboundq}
(they are also \emph{perfect} over a binary alphabet, and in some special cases
over a ternary alphabet \cite{kovacevic+vukobratovic_desi}).

\begin{corollary}
\label{thm:asympperfectq}
  The following statements hold for multiset codes over a $ q $-ary alphabet,
in the limit $ n \to \infty $.
\begin{itemize}
\item
For $ q = 2 $ and any $ h \geq 1 $,
\begin{align}
  M_{2}(n, h)  \sim  M^\textsc{l}_{2}(n, h)  \sim  \frac{ n }{ h + 1 }  .
\end{align}
\item
For $ q = 3 $ and any $ r \geq 1 $,
\begin{align}
  &M_{3}(n, 2r)    \sim  M^\textsc{l}_{3}(n, 2r)    \sim  \frac{ n^2 }{ 2 (3 r^2 + 3 r + 1) }  ,\\
	&M_{3}(n, 2r-1)  \sim  M^\textsc{l}_{3}(n, 2r-1)  \sim  \frac{ n^2 }{ 6 r^2 }  .
\end{align}
\item
For any $ q \geq 2 $ and $ h = 1 $,
\begin{align}
  M_{q}(n,1)  \sim  M^\textsc{l}_{q}(n,1)  \sim  \frac{ n^{q-1} }{ q! }  .
\end{align}
\item
For any $ q \geq 2 $ with $ {q-1} $ a prime power, and $ h = 2 $,
\begin{align}
  M_{q}(n,2)  \sim  M^\textsc{l}_{q}(n,2)  \sim  \frac{ n^{q-1} }{ ({q-1})!\ (q^2 - q + 1) }  .
\end{align}
\end{itemize}
\end{corollary}
\begin{IEEEproof}
  In all the stated cases there exist lattice tilings
$ \left( S_{q-1}\left(\lceil \frac{h}{2} \rceil, \lfloor \frac{h}{2} \rfloor\right) ,  \Lat \right) $
of $ \mathbb{Z}^{q-1} $, implying that $ \phi(h,q) = \beta(h,q) $
(these are the diameter-perfect codes of Theorems \ref{thm:perfect} and \ref{thm:diamperfect}).
The claim then follows from Theorem \ref{thm:asympboundsq} after plugging in the
expressions for $ \beta(h,q) $ in these particular cases.
\end{IEEEproof}

\subsection{Bounds and asymptotics: Growing alphabet case}
\label{sec:boundsqgrows}

We now discuss the case when the input alphabet is not necessarily fixed.
In particular, we consider the regime when the size of the alphabet is a linear
function of the block-length $ n $, namely $ q = \floor{\tilde{q} n} $ for an
arbitrary positive real constant $ \tilde{q} $.
As we shall point out in Section \ref{sec:construction2}, this regime is well-motivated
by the standard way of dealing with the effect of symbol reordering in networking
applications.

The upper bound in \eqref{eq:upperboundq} is not useful in this regime, so we derive
in Theorem \ref{thm:boundsqgrows} another bound appropriate for this case.
The method we use is similar to \cite{levenshtein}, though the setting is quite different.
Before stating the theorem, we give two auxiliary facts that will be needed in the proof.

\begin{lemma}
\label{thm:nonzero}
  Let $ {\bf x} \in \triangle^{q-1}_n $ be a vector with $ i $ non-zero coordinates.
The set of all vectors that can be obtained after $ \bf x $ is impaired with $ r $
deletions and $ h - r $ insertions%
\footnote{Recall that we are always referring to deletions and insertions in
the \emph{multisets} represented by vectors from $ \triangle^{q-1}_n $, not in
these vectors themselves.}
has at least $ \binom{i}{r} \binom{q-1+h-2r}{h-r} $ elements.
\end{lemma}
\begin{IEEEproof}
Subtract $ 1 $ from $ r $ of the positive coordinates of $ \bf x $, and distribute
a mass of $ h - r $ over the remaining $ q - r $ coordinates.
The former can be done in $ \binom{i}{r} $ ways, and the latter in $ \binom{q-1+h-2r}{h-r} $.
\end{IEEEproof}

\vspace{2mm}

There are $ \binom{q}{i} \binom{n-1}{i-1} $ vectors in $ \triangle^{q-1}_n $ with
exactly $ i $ non-zero coordinates.
Consequently,
\begin{equation}
\label{eq:allmultisets}
  \sum_{i=1}^{\min\{q, n\}} \binom{q}{i} \binom{n-1}{i-1} = |\triangle_n^{q-1}| = \binom{n+q-1}{q-1} .
\end{equation}

\begin{theorem}
\label{thm:boundsqgrows}
  Fix $ q \geq 2 $ and $ n > h \geq 1 $.
The following inequality is valid for all integers $ r \in \{0, 1, \ldots, h\} $ and
$ l \in \{r	, \ldots, q \} $:
\begin{align}
\label{eq:upperboundqgrows1}
	M_{q}(n,h)  \leq  \frac{ \binom{n + h - 2r + q-1}{q-1} }{ \binom{l}{r} \binom{q-1+h-2r}{h-r} }  +  
	                    \sum_{i=1}^{l-1}  \binom{q}{i} \binom{n-1}{i-1} .
\end{align}
In particular, for $ r = 0 $, $ l = 1 $, this simplifies to
\begin{align}
\label{eq:upperboundqgrows11}
	M_{q}(n,h)  \leq  \frac{ \binom{n + h + q-1}{q-1} }{ \binom{q-1 + h}{h} } .
\end{align}
\end{theorem}
\begin{IEEEproof}
  Let $ \C \subseteq \triangle^{q-1}_n $ be an optimal multiset code correcting $ h $
deletions, i.e., $ |\C| = M_{q}(n,h) $.
Write $ M_{q}(n,h) = M' + M'' $, where $ M' $ is the number of codewords of $ \C $
having at least $ l $ non-zero coordinates, and $ M'' $ the number of remaining codewords.
Recall from Theorem \ref{thm:indelsub} that $ \C $ corrects $ h $ deletions if and only
if it corrects $ r $ deletions and $ h - r $ insertions.
This implies that sets of outputs obtained by deleting $ r $ and inserting $ h - r $
symbols to each of the codewords, are disjoint.
Note that these outputs live in $ \triangle_{n+h-2r}^{q-1} $, which, together with Lemma
\ref{thm:nonzero}, implies that
$ M' \cdot \binom{l}{r} \binom{q-1+h-2r}{h-r} \leq \big| \triangle_{n+h-2r}^{q-1} \big| = \binom{n + h-2r + q-1}{q-1}  $.
This gives the first summand in the upper bound \eqref{eq:upperboundqgrows1}.
The second summand is simply the cardinality of the set of all vectors in $ \triangle^{q-1}_n $
having less than $ l $ non-zero coordinates, which is certainly an upper bound on $ M'' $.
\end{IEEEproof}

\vspace{2mm}

For a real $ \tilde{q} > 0 $ and an integer $ h \geq 1 $, define
\begin{equation}
\label{eq:chq}
  c(h, \tilde{q}) = \min_{0 \leq r \leq h} (h-r)!\, r!\, (1 + \tilde{q})^{r} .
\end{equation}

\begin{theorem}
\label{thm:boundsasympqgrows}
  For any fixed real $ \tilde{q} > 0 $ and integer $ h \geq 1 $, as $ n \to \infty $,
\begin{equation}
\label{eq:boundsasympqgrows}
  \frac{ \binom{ n + \floor{\tilde{q} n}-1 }{ \floor{\tilde{q} n}-1 } }{ \tilde{q}^h n^h }
   \lesssim   M_{\floor{\tilde{q} n}}(n,h)   \lesssim
    c(h, \tilde{q}) \frac{ \binom{ n + \floor{\tilde{q} n}-1 }{ \floor{\tilde{q} n}-1 } }{ \tilde{q}^h n^h } ,
\end{equation}
where
\begin{equation}
\label{eq:binomexp}
  \binom{ n + \floor{\tilde{q} n}-1 }{ \floor{\tilde{q} n}-1 }   \sim
	  \frac{ 2^{n (1 + \tilde{q}) H\left(\frac{1}{1 + \tilde{q}}\right)} }{ \sqrt{2 \pi  n} } 
				\left( 1 + \tilde{q}^{-1} \right)^{ r(n) - \frac{1}{2} } ,
\end{equation}
$ r(n) \defeq \floor{\tilde{q} n} - \tilde{q} n $, and
$ H(\cdot) $ is the binary entropy function.
\end{theorem}
\begin{IEEEproof}
  The expression \eqref{eq:binomexp} follows from the Stirling's approximation for the factorial.

The lower bound in \eqref{eq:boundsasympqgrows} follows from \eqref{eq:lowerbound}
and the Bose--Chowla construction of $ B_h $ sets \cite{bose+chowla} which implies that
$ \phi(h,q) < q^h + o(q^h) $ for fixed $ h $ and $ q \to \infty $, i.e., that
$ \phi(h,q) \lesssim q^h $.

The upper bound in \eqref{eq:boundsasympqgrows} is obtained from \eqref{eq:upperboundqgrows1}
after choosing $ l $ appropriately.
The idea is to set $ l $ large enough so that the first summand on the right-hand side of
\eqref{eq:upperboundqgrows1} is minimized, but small enough so that the second summand is
still negligible compared to the first one.
Observe the relation \eqref{eq:allmultisets} and the second summand on the right-hand side
of \eqref{eq:upperboundqgrows1}.
From \eqref{eq:allmultisets} we see that, as $ n \to \infty $ and $ q \sim \tilde{q} n $,
the sum $ |\triangle_n^{q-1}| = \sum_{i} \binom{q}{i} \binom{n-1}{i-1} $ grows exponentially in $ n $
with exponent $ (1 + \tilde{q}) H\big(\frac{1}{1 + \tilde{q}}\big) $ (see \eqref{eq:binomexp}),
and since it has linearly many summands, there must exist $ \lambda \in (0,1) $
such that $ \binom{q}{\lambda n} \binom{n-1}{\lambda n} $ grows exponentially
in $ n $ with the same exponent.
By using Stirling's approximation, one can find the exponent of $ \binom{q}{\lambda n} \binom{n-1}{\lambda n} $
as a function of $ \lambda $, and check by differentiation that it is maximized for
unique $ \lambda = \lambda^* = \frac{\tilde{q}}{1 + \tilde{q}} $.
Informally speaking, the term $ \binom{q}{\lambda^* n} \binom{n-1}{\lambda^* n} $
in the sum \eqref{eq:allmultisets}, and the terms ``immediately around it'', account
for most of the space $ \triangle^{q-1}_n $, and the remaining terms are negligible.
More precisely, there exists a sublinear function $ f(n) = o(n) $ such that
$ \sum_{i=1}^{\lambda^*n - f(n)} \binom{q}{i} \binom{n-1}{i-1} = o\big(n^{-h} \binom{n+q-1}{q-1}\big) $
\cite{spencer}.
Therefore, if we set $ l = \lambda^*n - f(n) $, the second summand on the right-hand side
of \eqref{eq:upperboundqgrows1} will be asymptotically negligible compared to the first
one, and the first summand will give precisely the upper bound in \eqref{eq:boundsasympqgrows}.
\end{IEEEproof}

\vspace{2mm}

In particular, the asymptotic expression for the cardinality of optimal
single-deletion-correcting multiset codes ($ h = 1 $) has the following form:

\begin{corollary}
\label{thm:qgrowsh1}
  For any positive real constant $ \tilde{q} $, as $ n \to \infty $,
\begin{equation}
  M_{\floor{\tilde{q} n}}(n, 1)  \sim  \frac{ 2^{n (1 + \tilde{q}) H\left(\frac{1}{1 + \tilde{q}}\right)} }
	                                  { \sqrt{2 \pi } \ n^{\frac{3}{2}} }
																		  \tilde{q}^{-1}  \left( 1 + \tilde{q}^{-1} \right)^{ r(n) - \frac{1}{2} } ,
\end{equation}
where $ r(n) \defeq \floor{\tilde{q} n} - \tilde{q} n $.
\end{corollary}
\begin{IEEEproof}
  Follows from the bounds in \eqref{eq:boundsasympqgrows}, and the fact that
$ c(1, \tilde{q}) = 1 $.
\end{IEEEproof}

\vspace{2mm}

On the logarithmic scale, we obtain from Theorem \ref{thm:boundsasympqgrows}:
\begin{align}
\nonumber
  \frac{1}{n} &\log_2 M_{\floor{\tilde{q} n}}(n,h)  \\
  \label{eq:redundancy2}
	& = (1 + \tilde{q}) H\Big(\frac{1}{1 + \tilde{q}}\Big) -
   \Big(h + \frac{1}{2}\Big) \frac{\log_2 n}{n}  +  {\mathcal O}\Big(\frac{1}{n}\Big) .
\end{align}
Hence, the largest rate achievable asymptotically
by multiset codes of block-length $ n $ over an alphabet of size $ \floor{\tilde{q} n} $
is $ (1 + \tilde{q}) H\big(\frac{1}{1 + \tilde{q}}\big) $ bits per symbol.
This is the capacity of the \emph{noiseless} multiset channel described in Section
\ref{sec:channel}.
The back-off from capacity at finite block-lengths scales as
$ \frac{1}{2} \frac{\log_2 n}{n}  +  {\mathcal O}\big(\frac{1}{n}\big) $ in the
noiseless case, while an additional redundancy of $ h \frac{\log_2 n}{n} $ is
necessary if the code is required to have the capability of correcting $ h $ deletions.
Note that the cardinality of codes produced by the construction based on Sidon
sets has the same asymptotic expansion \eqref{eq:redundancy2}.
In other words, the constructed codes are optimal in the sense of minimal asymptotic
code redundancy, for any $ h $ and any $ \tilde{q} $.

\subsection{Connections to classical binary insertion/deletion channels}
\label{sec:indel}

Codes in the simplex $ \triangle^{q-1}_n $ are relevant not only for permutation
channels and unordered data storage, but also for classical binary channels.
We shall omit the detailed description of this connection, as it can be found
in the relevant references;
instead, we only provide a brief comparison between the results in these references
and the results obtained here.

\vspace{2mm}

\subsubsection{Deletion and repetition channel with constrained inputs}

The classical binary deletion channel with inputs constrained in such a way that
all of them have the same number of runs of identical consecutive symbols, and
that each run is of length at least $ r $, can be reduced to the multiset channel
treated here via run-length coding \cite{sok, belfiore}.
More precisely, codes correcting deletions and repetitions of binary symbols
can be equivalently described in the metric space $ (\triangle^{q-1}_n, d_1) $,
for appropriately defined $ n $ and $ q $ (these parameters have an entirely
different meaning in this setting from that in ours).
In \cite{sok, belfiore}, the authors provide constructions and derive bounds on
the cardinality of optimal codes in $ (\triangle^{q-1}_n, d_1) $ having a specified
minimum distance.
Furthermore, the asymptotic regimes studied there correspond exactly to those
we discussed in the previous two subsections.

We wish to point out that the bounds obtained in this paper, Theorems \ref{thm:asympboundsq}
and \ref{thm:boundsasympqgrows} in particular, are strictly better that the ones in
\cite{sok, belfiore}.
The main reason for this is that the authors in \cite{sok, belfiore} derive bounds
on codes in $ \triangle^{q-1}_n $ under $ \ell_1 $ distance via packings in $ \mathbb{Z}^{q-1} $
under $ \ell_1 $ distance.
However, as we have shown in Theorem \ref{thm:isometry}, the $ \ell_1 $ distance
in the simplex corresponds to a different metric in $ \mathbb{Z}^{q-1} $, namely $ \da $.
The latter observation, together with the fact that optimal linear codes in $ (\mathbb{Z}^{q-1}, \da) $
can be constructed via Sidon sets (Theorem \ref{thm:Bh}), enables one to derive
better bounds.

\vspace{2mm}
\subsubsection{Channel with deletions of zeros}

Restricting the inputs of a binary channel to sequences of the same Hamming weight,
and describing such sequences by their \emph{runs of zeros}, one again obtains
$ \triangle^{q-1}_n $ as the relevant code space.
This representation of binary sequences is appropriate for the binary deletion
channel in which only zeros can be deleted%
\footnote{We also note that channels with deletions of zeros are equivalent to
channels with duplication errors; see, e.g., \cite{dolecek,jain}.};
see \cite{levenshtein}.
While \cite{levenshtein} does not discuss the constant-weight case but rather the
binary channel with no constraints on inputs, the methods of analysis are similar,
at least in some asymptotic regimes.
For example, the construction via Sidon sets was given also in \cite{levenshtein}
and, when appropriately modified for the constant-weight case, implies a lower
bound similar to the one stated in Theorem \ref{thm:boundsasympqgrows}.
(Levenshtein was unaware of \cite{bose+chowla} and the construction of Sidon sets
therein.
Consequently, he stated in \cite{levenshtein} an explicit lower bound which is
worse than what his construction actually implies.)

It is worth noting that the upper bound we have derived in the previous subsection
improves on that in \cite{levenshtein}.
In particular, there is no need to distinguish between the cases of odd and even
$ h $, as was done in \cite{levenshtein}.

\section{Other constructions of multiset codes}
\label{sec:otherconstr}

In this section we describe two additional constructions of multiset codes.
Both of these constructions are asymptotically suboptimal and result in codes
of smaller cardinality compared to the construction based on Sidon sets, but
are of interest nonetheless.

\subsection{Construction based on sequence number prefixes}
\label{sec:construction2}

In networking applications, particularly those employing multipath routing, the
problem of packet reordering is usually solved by supplying each packet with a
sequence number in its header \cite{kurose+ross}.
If up to $ n $ packets are being sent in one ``generation'', a sequence number
will take up $ \lceil \log_2 n \rceil $ bits.
Therefore, if the original packets are of length $ l $ bits each, i.e., the cardinality
of the channel alphabet is $ \tilde{q} = 2^l $, then the ``new'' packets with
prepended sequence numbers will be of length $ l + \lceil \log_2 n \rceil $ bits.
Notice that by adding sequence number prefixes, we are actually changing the channel
alphabet---the new alphabet is the set of all packets of length $ l + \lceil \log_2 n \rceil $,
and its cardinality is $ 2^{l + \lceil \log_2 n \rceil} \approx \tilde{q} n $.

Furthermore, in order to protect the packets from other types of noise, a classical
code of length $ n $ in the $ \tilde{q} $-ary Hamming space may be used \cite{kovacevic+vukobratovic_clet}.
To clarify what is meant here, we are assuming that:
\begin{inparaenum}
\item
a sequence of information packets to be transmitted, $ (s_1, \ldots, s_n) $, is a codeword
of a code $ \C_\textsc{h} $ of length $ n $ over a $ \tilde{q} $-ary alphabet having minimum
Hamming distance $ > h $, and
\item
to each symbol/packet of this codeword we then prepend a sequence number indicating its
position in the codeword, i.e., the sequence actually transmitted is $ (u_1, \ldots, u_n) $,
where $ u_i = i \circ s_i $ (`$ \circ $' denotes concatenation).
\end{inparaenum}
Notice that the order in which $ u_i $'s are transmitted is irrelevant because it can
easily be recovered from the sequence numbers.
In other words, each codeword $ (u_1, \ldots, u_n) $ obtained in the above-described
way can be thought of as a multiset $ \ldblbrace u_1, \ldots, u_n \rdblbrace $.
Therefore, the resulting code $ \C $ can in fact be seen as a multiset code over an
alphabet of size $ \tilde{q} n $, but a special case thereof in which no codeword
contains two identical packets (as each of the $ n $ packets has a different sequence
number prefix).
Furthermore, the fact that $ \C_\textsc{h} $ has minimum Hamming distance $ > h $
implies that $ \C $ can correct $ h $ packet deletions.

To compare this construction with the one given in Section \ref{sec:construction},
note that the size of the optimal code that can be obtained in this way, denoted
$ M^\text{seq}_{\tilde{q} n}(n,h) $, cannot exceed the sphere packing bound in the
$ \tilde{q} $-ary Hamming space:
\begin{equation}
  M^\text{seq}_{\tilde{q} n}(n,h)  \leq
     \frac{ \tilde{q}^n }{ \sum_{j=0}^{\lfloor \frac{h}{2} \rfloor} \binom{n}{j}(\tilde{q}-1)^j }  \sim
     \frac{ \lfloor \frac{h}{2} \rfloor ! \ \tilde{q}^n }
          { (\tilde{q}-1)^{\lfloor \frac{h}{2} \rfloor} \ n^{\lfloor \frac{h}{2} \rfloor} } .
\end{equation}
From Theorem \ref{thm:boundsasympqgrows} we then get, for any fixed $ \tilde{q} \geq 2 $
and $ h \geq 1 $, as $ n \to \infty $,
\begin{equation}
 \begin{aligned}
   \frac{ M_{\tilde{q} n}(n,h) }{ M^\text{seq}_{\tilde{q} n}(n,h) }
     \gtrsim  2^{ n (1+\tilde{q}) \log_2(1 + \tilde{q}^{-1})  + o(n) }
 \end{aligned}
\end{equation}
and hence
%\begin{equation}
 \begin{align}
\nonumber
   \frac{1}{n} \log_2 M_{\tilde{q} n}(n,h)  - 
	  \frac{1}{n} \log_2 &M^\text{seq}_{\tilde{q} n}(n,h)   \\
    &\gtrsim (1+\tilde{q}) \log_2(1 + \tilde{q}^{-1}) .
 \end{align}
%\end{equation}
In words, the asymptotic rate achievable by multiset codes based on sequence number
prefixes is strictly smaller than the corresponding rate achievable by general multiset
codes.
The lower bound on the difference, $ (1+\tilde{q}) \log_2(1 + \tilde{q}^{-1}) $, is
a monotonically decreasing function of $ \tilde{q} $, and hence also of the length of
information packets $ l = \log_2 \tilde{q} $.
Thus, the savings (in terms of rate) obtained by using optimal multiset codes instead
of the ones based on sequence numbers are greater for large block-lengths and small
alphabets.

\begin{remark}
\textnormal{
  It should be noted that, in some other contexts and asymptotic regimes, the construction
based on the sequence number prefixes is optimal in terms of achievable rates; see \cite{heckel}.
}
\myqed
\end{remark}

\subsection{Construction based on polynomial roots}
\label{sec:construction3}

Consider a polynomial
\begin{equation}
\label{eq:polynomial}
  s(x) = x^n + s_{n-1} x^{n-1} + \cdots + s_1 x + s_0 ,
\end{equation}
with coefficients $ s_i $ drawn from a finite field $ \mathbb{F}_{p^m} $
($ p \geq 2 $ is a prime and $ m \geq 1 $ an integer).
Each such polynomial has $ n $ (not necessarily distinct) roots, $ u_1, \ldots, u_n $,
which are elements of the extended field $ \mathbb{F}_{p^{mn}} $.
The coefficients $ (s_0, s_1, \ldots, s_{n-1}) $ can always be recovered from the
roots, e.g., by using Vieta's formulas:
\begin{equation}
\label{eq:vieta}
  s_{n-k} =  (-1)^k \sum_{1\leq i_1 < \cdots < i_k \leq n}  u_{i_1} \cdots u_{i_k}  ,
\end{equation}
for $ k = 1, \ldots, n $.
Therefore, \emph{multisets} of roots $ \ldblbrace u_1, \ldots, u_n \rdblbrace $ are
in one-to-one correspondence with sequences of coefficients $ (s_0, s_1, \ldots, s_{n-1}) $,
and the mapping:
\[ (s_0, s_1, \ldots, s_{n-1}) \mapsto \ldblbrace u_1, \ldots, u_n \rdblbrace \]
defines a multiset code of length $ n $ over an alphabet of size $ p^{mn} $.
In channel coding parlance, $ (s_0, s_1, \ldots, s_{n-1}) $ is an information sequence
and $ \ldblbrace u_1, \ldots, u_n \rdblbrace $ is the corresponding codeword to be
transmitted.

Furthermore, we can extend this construction to obtain a code capable of correcting
$ h \geq 1 $ deletions, which in the present terminology means that the coefficients
of the ``information polynomial'' \eqref{eq:polynomial} can be recovered from any
$ n - h $ of the $ n $ transmitted roots.
To do this, fix $ h $ coefficients beforehand, say $ s_{n-1} = \cdots = s_{n-h} = 0 $.
We see from \eqref{eq:vieta} that the remaining coefficients can indeed be recovered
uniquely from any $ n - h $ roots.
In other words, the information sequence is now $ (s_0, s_1, \ldots, s_{n-h-1}) $, and
the mapping:
\[ (s_0, s_1, \ldots, s_{n-h-1}, 0, \ldots, 0)   \mapsto
   \ldblbrace u_1, \ldots, u_n \rdblbrace , \quad s_i \in \mathbb{F}_{p^m} , \]
defines a multiset code with the following properties:
length $ n $, minimum distance $ > h $, and cardinality $ p^{m(n-h)} $.
The code is defined over an alphabet of size $ p^{mn} $ ($ u_i \in \mathbb{F}_{p^{mn}} $).

\section{Concluding remarks and further work}
\label{sec:conclusion}

We have described a coding-theoretic framework for a communication setting where
information is being transmitted in the form of multisets over a given finite alphabet.
General statements about the error correction capability of multiset codes have
been obtained, constructions of such codes described, and bounds on the size of
optimal codes derived.
Furthermore, the \emph{exact} asymptotic behavior of the cardinality of optimal
codes has been obtained in various cases.

As we have shown, the study of multiset codes over a fixed alphabet reduces to
the study of codes in $ A_m $ lattices, at least in the large block-length limit.
In connection to this, there are several natural directions of further work
on this topic:
\begin{itemize}
\item
Improve the bounds on the density of optimal codes in $ (A_m, d_1) $ having a given
minimum distance.
\item
Investigate whether the construction of codes in $ (A_m, d_1) $ via Sidon sets is optimal,
or it is possible to achieve larger densities with non-linear codes.
\item
Demonstrate (non-)existence of diameter-perfect codes in $ (A_m, d_1) $ with parameters
different from those stated in Theorems \ref{thm:perfect} and \ref{thm:diamperfect}.
\end{itemize}

We have also argued that the case where the alphabet size is a growing function
of the block-length is a meaningful asymptotic regime for multiset codes.
Several problems worth investigating in this context are:
\begin{itemize}
\item
Improve the bounds on $ M_{\tilde{q} n}(n; h) $ for $ h \geq 2 $.
\item
Investigate whether the construction of multiset codes via Sidon sets is optimal in
this regime.
\item
Derive bounds on $ M_{\tilde{q} n}\big(n; \tilde{h} n\big) $, for an arbitrary
constant $ \tilde{h} \in (0,1) $.
Namely, the number of deletions growing linearly with the block-length is another
natural asymptotic regime.
Note that the upper bound in \eqref{eq:upperboundqgrows11} is valid in this regime
and can be expressed in the corresponding asymptotic form via \eqref{eq:binomexp}.
As for the lower bound, one can apply the familiar Gilbert--Varshamov bound.
However, due to the structure of the space $ \triangle_n^{q-1} $ and, in particular,
the fact that balls in this space do not have uniform sizes, we do not expect this
bound to be tight.
\end{itemize}

Finally, other approaches to error correction in permutation channels and unordered
data storage systems may be more appropriate in some settings than the one presented
here.
For example, rather than regarding errors at the DNA molecule level simply as
substitutions of molecules, one might consider applying error correcting codes
both at the molecule level and over all the molecules stored in a pool.
Also, more generally, it would be worthwhile investigating problems other than
error correction that are related to communication and data storage in this context%
\footnote{See, e.g., \cite{steinruecken, varshney+goyal} for a discussion of compression
problems related to (multi)sets.}.
The ``unordered'' kind of information transfer has a clear disadvantage of reducing
the achievable rates significantly, but, as discussed in Section \ref{sec:motivation},
there may be situations where it is necessary, either due to the inherent characteristics
of the communication channel, or the technological constraints of the writing/reading
process in a data storage system.

\vspace{2mm}
\section*{Acknowledgment}

We thank Dr. Aslan Tchamkerten and Dr. Jossy Sayir for sending us the preprints
\cite{belfiore} and \cite{mackay}, respectively;
Prof. David Tse and Dr. Netanel Raviv for helpful discussions on the subject of
DNA storage;
and the anonymous referees for useful comments on the original version of the
manuscript.

\IEEEtriggeratref{38}

\end{document}